# Selection and Behavioral Responses of Health Insurance Subsidies in the Long Run: Evidence from a Field Experiment in Ghana*

Patrick Opoku Asuming, Hyuncheol Bryant Kim, and Armand Sim

April 2021

**Abstract**

We conduct a randomized experiment that varies one-time health insurance subsidy amounts (partial and full) in Ghana to study the impacts of subsidies on insurance enrollment and health care utilization. We find that both partial and full subsidies promote insurance enrollment in the long run, even after the subsidies expired. Although the long run enrollment rate and selective enrollment do not differ by subsidy level, long run health care utilization increased only for the partial subsidy group. We show that this can plausibly be explained by stronger learning-through-experience behavior in the partial than in the full subsidy group.

---

* Asuming: University of Ghana Business School (email: poa2102@gmail.com); Kim (corresponding author): Department of Economics, Hong Kong University of Science and Technology and Department of Policy Analysis and Management, Cornell University (email: hk788@cornell.edu); Sim: Centre for Development Economics and Sustainability, Monash University (email: armand.sim@monash.edu). We thank Ama Baafra Abeberese, Douglas Almond, Diane Alexander, Jim Berry, John Cawley, Esteban Mendez Chacon, Pierre-Andre Chiappori, Giacomo de Giorgi, Junlong Feng, Amy Finkelstein, Rema Hanna, Supreet Kaur, Robert Kaestner, Don Kenkel, Daeho Kim, Michael Kremer, Wojciech Kopczuk, Leigh Linden, Corrine Low, Doug Miller, Sangyoon Park, Seollee Park, Cristian Pop-Eleches, Bernard Salanie, and seminar participants at Columbia University, Cornell University, Seoul National University, and the NEUDC. This research was supported by the Cornell Population Center and Social Enterprise Development Foundation, Ghana (SEND-Ghana)." Armand Sim gratefully acknowledges financial support from the Indonesia Education Endowment Fund. All errors are our own.

## 1. Introduction

Many developing countries have increasingly been instituting social health insurance schemes (SHIs) to help mitigate the effects of adverse health shocks, especially for the poor (WHO, 2005, 2010).[1] The SHIs offer low sign-up costs and generous benefits, but the take-up and retention rates remain low in many countries (Fenny et al., 2016), especially among the poorest households (Acharya et al., 2013).

Achieving universal coverage or a high enrollment rate is important in terms of risk pooling and the sustainability of SHIs. To ease financial burden and sustain SHIs, some countries, such as Ghana and Indonesia introduced a contributory system with individual mandates (Banerjee et al, 2020). Enforcing mandates in this system is important because otherwise the insurance can become financially unsustainable if people who are more ill or use more health care service selectively enroll in social health insurance. However, due to administrative constraints the governments in developing countries find it difficult to successfully impose mandates.

Recent studies find that various efforts to promote health insurance enrollment and health care utilization have limited impact (Thornton et al., 2010; Capuno et al., 2016; Wagstaff et al., 2016).[2] Subsidies are one of the few successful types of interventions used to promote health insurance enrollment in developing countries (e.g., Thornton et al., 2010; Banerjee et al., 2020). However, there are two important aspects of the effects of subsidy intervention that remain relatively less understood. The first aspect is the effects of subsidy level (price). Different levels of subsidies may attract people with different characteristics, and this selection may affect the health care service utilization among the insured.[3] The screening effect of subsidy level has been studied for a few health products and services, but it has not been intensively investigated for health insurance in a developing country setting, with a few exceptions of Banerjee et al (2020) on Indonesia and Kinnan et al (2020) on India.[4]

[1] Recent examples of countries that have instituted SHIs include Ghana, Kenya, Nigeria, Tanzania, and Vietnam. Countries in the process of instituting SHIs include Cambodia, Laos, Malaysia, Zimbabwe, and South Africa. (Wagstaff, 2010).
[2] Wagstaff et al. (2016) and Capuno et al. (2016) find subsidy and information do not successfully promote health insurance enrollment. Thornton et al. (2010) find subsidy increases short-term enrollment but does not increase health care service utilization.
[3] In addition, as Dupas (2014) explains, the price level may affect the long-run adoption of health products through the "anchoring" mechanism, where a previously encountered price may act as anchor and affect people's valuation of a product regardless of its intrinsic value.
[4] Some examples include facility delivery (Grepin et al, 2019), bed nets (Cohen and Dupas, 2010), and chlorine for water purification (Ashraf, Berry, and Shapiro, 2010).

The second aspect is the effects of receiving a temporary subsidy on long-run insurance enrollment and its consequences. The long-run effects are potentially ambiguous in theory. A temporary subsidy can reduce take-up in the long run if people consider the temporarily subsidized price as the reference point that could affect their future reservation price (Simonson and Tversky, 1992; Koszegi and Rabin, 2006). On the other hand, a temporary subsidy can increase demand in the long run for an experience good, i.e., health insurance, if people using the subsidized good subsequently learn the value and the benefits of it. Despite the importance of understanding the long-run effects of temporary subsidy, there is relatively scarce evidence on its application and evaluation on long-run health insurance enrollment in developing countries, except Banerjee et al (2020).

We complement the literature by studying the long-run impacts of randomly providing different levels of one-time health insurance subsidies on enrollment, health care utilization, and health outcomes. Specifically, we randomly selected communities for the subsidy intervention and further randomized different levels of subsidies—partial (one-third and two-thirds) and full—for the insurance premiums and fees for one-year of coverage at the household level. To measure the impact of these interventions, we conducted a baseline survey and two follow-up surveys at seven months and at three years after the initial intervention.

We study selection in enrollment by subsidy level. Specifically, we examine selection into insurance by investigating whether the subsidy induces those who are *ex-ante* more ill and use more health care services to enroll in health insurance. In this case, selection exists if compliers—those who self-select to enroll in health insurance in response to a partial or full subsidy—have greater *ex-ante* health risk or use more health care services than never takers—those who do not enroll in health insurance regardless of the subsidy. In theory, selection could be stronger in the partial subsidy group than that of the full subsidy group because those who have to pay an insurance premium might enroll more selectively than those who do not have to.

In addition to *ex-ante* selection, we also evaluate *ex-post* behavioral responses by investigating whether enrollees that pay higher insurance prices are more likely to utilize more health care services (Einav and Finkelstein, 2011; Chiappori and Salani´e, 2013). We propose two possible channels that can explain why *ex-post* behavioral responses may arise in our context. The first channel is *learning-through-experience*, which occurs when enrollees that have better access to health care services become more familiar with them. Because individuals may not have learned the insurance benefits until they experienced it, familiarity with health services may increase its usage, especially in the long run. Learning-through-experience might differ by level of subsidy. For example, those who pay insurance premium (enrolled in response to partial subsidy) could seek and learn about health care providers more actively than those who do not (enrolled in response to full subsidy) because people that must pay higher cost tend to reduce information asymmetry associated with health care providers (Gaynor, 1994; McGuire, 2000). This channel predicts that the impacts on health care utilization could be more apparent in the long run than in the short run because learning could be slow and imperfect.

The second channel is related with *sunk cost fallacy*, which states that an individual is more likely to use a product when he/she pays a price for it (Thaler, 1980; Eyster, 2002). In our context, it predicts that those in the partial subsidy group could feel that they must use health care services paid for by the NHIS because they pay some money for insurance premiums, not necessarily because they are ill. Because the subsidy covers premium in the short run, this channel predicts that the impacts of partial subsidy—compared to full subsidy—on health care utilization in the short run could be positive. This channel also predicts that short-run effects could be stronger than long run effects, which are measured after the subsidy expired.

We find a significant increase in short- and long-run insurance take-ups. Those receiving one-third, two-thirds, and a full subsidy were 36.9, 47.6, and 51.7 percentage points, respectively, more likely to enroll in health insurance in the short-run. Three years after the initial intervention, we still observe increased enrollment. Those who received one-third, two-thirds, and a full subsidy were 17.9, 14.3, and 20.4 percentage points, respectively, more likely to enroll in health insurance in the long run.

We find evidence of selection by subsidy. In general, those who enrolled in response to the subsidy (compliers) are more ill and utilize more health care services at the baseline than those who did not enroll regardless of the subsidy (never-takers). However, we do not find evidence that selection vary by subsidy level (partial vs. full subsidy). Interestingly, we find that health care services utilization in the long run increases only for the partial subsidy group, but not for the full subsidy group. We provide evidence that it is because *learning-through-experience* effect in the partial subsidy group is stronger than that in the full subsidy group.

In summary, this study shows that a short-term subsidy intervention can successfully sustain health insurance enrollment. We observe selection by subsidy in terms of *ex-ante* health risk, but we do not observe differences in selection between the partial and full subsidy. We also observe *ex-post* behavioral response (increased health care utilization) among the partial subsidy group, but not the full subsidy group. Together, these results suggest that subsidy, especially partial subsidy, could negatively affect the financial sustainability of social health insurance program.

Our study contributes to three strands of literature. First, our study is related to the literature on the impacts of subsidy or price intervention on selection in health insurance in developing countries. This topic has been relatively under explored in the literature, with a few recent exceptions (Fischer et al, 2018; Banerjee et al, 2020; Kinnan et al, 2020). We contribute to this literature in two ways. First, we provide different levels of insurance subsidy on enrollment allowing us to study selection by level of subsidy. Specifically, we study whether the characteristics of people who enrolled in health insurance vary by level of subsidy. Second, our two rounds of follow-up surveys within three years of the intervention allow us to study the dynamics of selection, which is relatively rare in a developing country context.

Second, our study contributes to the literature on sustainability of health intervention programs. This study is, to our knowledge, one of the few to document evidence of the long-run effects of interventions on insurance enrollment and retention in a developing country.[5] While the idea of promoting sustainability is attractive, it is difficult to achieve in practice. The challenges in promoting sustainable health insurance enrollment could be even greater because health care services in developing countries are generally of low quality and unreliable.[6] Few studies on this topic find mixed results on the effects of a one-time subsidy. In contrast to Kremer and Miguel (2007), who find limited evidence that a subsidy promotes long-run adoption of worm treatment, Dupas (2014) finds that a one-time subsidy is effective in boosting long-run adoption of bed nets.[7]

Lastly, our study contributes to the broad empirical literature on the effects of health insurance coverage on health care utilization and health status, which has so far produced mixed evidence. Some studies find insignificant impacts of insurance coverage on health outcomes (Thornton et al. 2010; Fink et al., 2013; King et al., 2009), while others find positive impacts (e.g., Miller, Pinto, and Vera-Hernandez, 2013; Gruber, Hendren, and Townsend, 2014). In terms of OOP expenses, some studies observe no or adverse effects of insurance on such expenses (e.g., Thornton et al., 2010; Fink et al., 2013), while others find the opposite (e.g., Galárraga et al., 2010). Our study is among the first to examine the effects of insurance coverage on both short- and long-run health outcomes in a low-income setting, while the existing literature largely focuses on short-run outcomes.[8]

[5] One exception is Banerjee et al. (2020) that studies the longer-run (32 months post-intervention) effects of different level of temporary subsidy (half and full), assisted registration, and information on participation in health insurance program in urban areas in Indonesia.
[6] See, for example, Banerjee, Deaton, and Duflo, (2004), Goldstein, et al. (2013), and Das, et al., (2016) for illustrations of low health care quality in developing countries. Alhassan et al. (2016) provides illustrations for Ghana.
[7] It is important to note, however, that the long-run effect of a one-time health insurance intervention is quite different from that of health product adoption such as worm treatment and malaria bed nets. Having health insurance does not necessarily result in improved health status. To be successful, health insurance enrollment should promote health care service utilization and prevent moral hazard behaviors. In addition, learning about the effects of other health products, such as deworming medicine, bed nets, and water disinfectants, could be less setting-specific than the case of health insurance, where the quality of health care services could vary considerably across settings.

The remainder of this paper is structured as follows. Section 2 outlines the research context. Section 3 describes the experimental design and data. Section 4 presents the empirical strategy and Section 5 presents the main results. Section 6 concludes the paper.

## 2. Institutional Background

### 2.1. National Health Insurance Scheme (NHIS) in Ghana

The National Health Insurance Scheme (NHIS) in Ghana was established by the National Health Insurance Act (Act 560) in 2003. It aims to improve access to and the quality of basic health care services for all citizens, especially the poor and vulnerable (Ministry of Health, 2004). The law mandates that every citizen enroll in at least one scheme. However, in practice, there are no penalties for those who do not enroll. Most of the 170 administrative districts of Ghana operate their own District Mutual Health Insurance Scheme (DMHIS) (Gajate-Garrido and Owusua, 2013).[9] Each DMHIS accepts and processes applications, collects premiums (and fees), provides membership identification cards, and processes claims from accredited facilities for reimbursement.

[8] In the US setting, a RAND experiment reports insignificant effects of insurance coverage on average health outcomes but finds negative effects on health outcomes for the more vulnerable subgroups (Newhouse and the Insurance Experiment Group, 1993). More recent studies find positive effects of exposure to public health insurance during childhood on various long-term health outcomes (Boudreaux, Golberstein, and McAlpine, 2016; Currie, Decker, and Lin, 2008; Miller and Wherry, 2016).

[9] There are three types of insurance schemes in Ghana: District Mutual Health Insurance Schemes (DMHIS), Private Mutual Health Insurance Schemes (PMHIS), and Private Commercial Insurance Schemes (PCHIS). The focus of this study is DMHIS, which represents 96 percent of insurance coverage (GSS, GHS and ICF, 2009). They are operated and subsidized by the government through the National Health Insurance Fund (NHIF). PMHIS are non-profit non-subsidized schemes run by NGOs, religious bodies and cooperative societies. PCHISs are for profit schemes that do not receive government subsidies.

Annual means-tested premiums, which are charged to informal sector workers, range from $5 to $ 32. However, owing to the lack of information on household incomes, rural districts tend to charge the lowest premiums, while urban districts charge higher premiums. Indigents, pregnant women, children under 18 years, and the elderly over 70 years are exempt from premiums but not registration fees.[10] All members, except for indigents and pregnant women, are required to pay registration fees when they first register and when they renew. Those who do not renew their membership by the due date pay penalties when they eventually renew their memberships.

The benefits package of the NHIS, which is the same across DMHISs, is very generous, albeit new members wait for three months before they can enjoy the insurance benefits. As described in Table A1, the package covers: 1) full outpatient and inpatient (surgery and medical) treatments and services, 2) full payment for medications on the approved list, 3) payments for referrals on the approved list, and 4) all emergencies. The NHIA (2010) estimates that 95% of disease conditions that affect Ghanaians are covered by the scheme. Those who enroll do not pay deductibles or copayments for health care service utilization by law; however, according to the USAID (2016), health care providers often charge unauthorized fees that are inaccurately described as copayments.

Despite the low premiums and generous benefits, enrollment in the NHIS remains low. By the end of 2010, the total active membership stood at 34% of the population of Ghana (NHIA, 2011). Enrollment is particularly low among the poorest. A 2008 nationwide survey found that only 29% of the individuals in the lowest wealth quintile were active members of the scheme compared to 64% of households in the highest quintile (National Development Planning Commission, 2009).

In addition to the lack of affordability, negative perceptions of the NHIS explain the low enrollment rate. For example, Alhassan et al. (2016) note that those enrolled in the NHIS generally perceive they are not receiving good-quality health care, for reasons such as long wait times and the poor attitudes of health staff towards patients. Additionally, Fenny et al. (2016) observe that perceived quality of service and socio-cultural factors such as trust, bad attitudes of health facility staff, and drug shortages contribute to low enrollment and retention rates in Ghana.

[10] The law defines an indigent as “a person who has no visible or adequate means of income or who has nobody to support him or her and by the means test.” Specifically, an indigent is a person who satisfies all of these criteria: i) unemployed and has no visible source of income, ii) does not have a fixed place of residence according to standards determined by the scheme, iii) does not live with a person who is employed and who has a fixed place of residence, and iv) does not have any identifiably consistent support from another person.

### 2.2. Setting

This study was conducted in Wa West, a poor and remote rural district in Northern Ghana (Figure A1). It covers an area of approximately 5,899 $km^2$ and had a population of about 81,000 in 2010. Settlement patterns are highly dispersed, with most residents living in hamlets of about 100-200 people. This high dispersion, coupled with the poor road network, makes traveling within the district difficult and expensive. The economy is largely agrarian, with over 90% of the population working as farmers. Estimates from the 2006 Ghana Living Standard Survey indicate that average annual per-capita income and health expenditure in a rural savannah locality like Wa West were about $252 and $26, respectively (Ghana Statistical Service, 2008).

In the study area, even though the Community-Based Health and Planning Services (CHPS) has increased accessibility to health care services,[11] there are only six public health centers and no tertiary health facility. During the study period, the district had only 15 professional nurses and no medical doctor (Nang-Beifua, 2010). The district also has a high disease burden. The most common cause of outpatient visits in the region is malaria, which accounts for one third of outpatient visits. Other common causes of outpatient visits are acute respiratory-tract infections and skin diseases.

The Wa West DMIHS was introduced in January 2007. In 2011, it charged a uniform premium of $5.46 (GHC 8.20) for adults (18-69) and a processing fee of $2.67 (GHC 4) for first-time members and $0.60 (GHC 1) for renewals. Late renewals incur a fee of $1.30 (GHC 2) in addition to full premiums for all years for which membership was not renewed.[12] The baseline enrollment rate in 2011 for the study sample is 20%.

## 3. Research Design

In this section, we discuss the original study, data collection, definition and construction of key variables, descriptive statistics as well as the balance test of baseline characteristics.

[11] CHPS are community health facilities that provide primary health care. They are located within rural communities with limited access to larger hospitals and are manned by nurses. Among the services offered are treatment of common ailments (malaria and diarrheal diseases) and maternal and child care services.
[12] The exchange rate at the time of the study was $1 = GHC 1.5.

### 3.1. Interventions

We begin by discussing the original study aimed to analyze short-run outcomes (Asuming, 2013). Three different interventions were introduced to 4,406 individuals of 629 household in 59 communities: a subsidy for the insurance premiums and fees (*Subsidy*), an information campaign on the national health insurance (*Campaign*), and an option for individuals to sign up in their community instead of traveling to the district capital (*Convenience*). Interventions were overlapping and randomized at the community level. Figure B.1 summarizes our original research design.[13]

The *Subsidy* intervention was conducted in two stages. In the first stage, the subsidy was randomly provided to households across communities. In the second stage, the level of the subsidy was randomized at the household level within the *Subsidy* communities: one-third ($2.67), two-thirds ($5.40), or full ($8.13) subsidy (see Figure 1). Subsidies were given in the form of vouchers, which were distributed between November 2011 and January 2012, valid for two months, and redeemable at the Wa West DMHIS center.[14]

The subsidy voucher specified the names, ages, and genders of all household members, expiration date, and place of redemption. Households that did not receive a full subsidy were informed about the extra amount needed to register all members. Although a subsidy was provided at the household level, enrollment had to be specified for each household member. Households do not necessarily insure all members.

We extended the original study by implementing a long-term follow-up survey and focus only on the *Subsidy* intervention. To formally support our approach, we conducted two empirical exercises. First, complementarity tests between *Subsidy* and other treatments. Obtaining unbiased causal effects of the *Subsidy* intervention requires no complementarity between *Subsidy* and the other treatments. Appendix Table B.1. shows that eight complementarity tests (e.g., *Sub* + *Camp* = *Sub* & *Camp*) fail to reject the null hypothesis of no complementarity at the 5 % level.

Second, restricting the sample to the control and *Subsidy* only groups (i.e., excluding *Subsidy* + *Campaign*, *Subsidy* + *Convenience*, and *Subsidy* + *Campaign* + *Convenience*) to investigate the cleaner effects of subsidy variation. We provide the estimation results in Appendix Tables B.2 (effects on enrollment), B.3 and B.4 (short- and long-run effects on health care utilization), and B.5 and B.6 (short- and long-run effects on health outcomes). The interpretation of our main findings and conclusion holds. Overall, the results of these two exercises lend support for our approach.

[13] The initial intention of the original study was to analyze the effects of single intervention and complementarity among interventions. We also estimate the long-term effects of all intervention. Results are shown in the Appendix B.
[14] The voucher could also be used to either initiate or renew insurance membership. Those who did not enroll at the baseline (80 %) could use the voucher anytime. Those who had already enrolled at the baseline (20 %) could only use the voucher if their existing renewal was due within the voucher's validity period. Otherwise they could not use the voucher.

### 3.2. Data Collection

The study sample includes 2,954 individuals from 418 households in 44 communities. We conducted the baseline survey in September 2011 and implemented the intervention in October 2011. We conducted two follow-up surveys seven months and three years after the intervention. The baseline survey collected information on demographic characteristics, employment, health status, health care service utilization, enrollment in the NHIS, and health behaviors for all household members.

The first follow-up survey collected information on health care service utilization, health status, and health behaviors. In the second follow-up survey, we collect sets of information similar to those in the first follow-up survey but with greater detail to improve the quality of the data. For example, we asked for specific dates and the respondent's status since the first follow-up for up to three episodes of several important illnesses, such as malaria, acute respiratory diseases, and skin diseases. As a result, there are some differences in the construction of short- and long-run utilization measures that prevent a direct comparison of health care service utilization and health status in these survey periods.[15]

The main outcome variables of interest, measured at the individual level, are health insurance enrollment, health care service utilization, and self-reported health status. Health care service utilization is measured by health facility visits in the last four weeks and last six months as well as OOP health expenditure. Health status is measured by the number of days of illness in the last four weeks, the indicator and the number of days an individual was unable to perform normal daily activities due to illness, and self-rated health status.[16] The measure of inability to perform normal daily activities is essentially similar to the measure of Activities of Daily Living (ADL) that is commonly used in the literature as an objective measure for health status.[17]

[15] The health facility visit variable in the first follow-up survey is constructed from the following question: "The last time (in the last four weeks/last six months) (NAME) was ill or injured, did he/she visit any health facility?". In the second follow-up survey, the same variable is constructed from questions about respondents' visits during illness episodes. For example, an individual is said to visit a health facility in the last six months if his/her illness episode occurred in the last six months and he/she sought treatment in the health facility.

[16] Self-rated health status, which is restricted to those aged 18 years or older, is only available in the follow-up surveys.

The attrition rate in the first follow-up survey was relatively low (5 %) but increases in the second follow-up survey (21 %), as shown in Appendix Table A2.[18] The short- and long-run attrition rates are not systematically correlated with our interventions.

### 3.3. Baseline Characteristics and Balance Test

Table 1 presents the summary statistics of baseline characteristics and balance tests between the intervention and control groups. Panels A and B report mean of baseline control (socio-economic and community) and outcome variables. Columns 1 to 5 report mean of baseline characteristics of all respondents, control group, and treatment groups. The average respondent is about 24 years old and 48% are male. About 20% were enrolled in the NHIS at the baseline survey, and 36% had ever registered with the scheme. In terms of health characteristics, 12% reported a sickness or injury in the last four weeks, about 4% visited a health facility in the last month, and 14% made a positive OOP health expenditure. The average household lives within 5.4 km of a health facility and 20 km from the district capital.

Columns 3 to 5 present results from regressions of each variable on control and subsidy level indicators. Only 2 out of 66 coefficients across all balance tests are statistically significant at the 5% level, which is what we expect by chance. We also compare the baseline differences between each subsidy level group in Columns 7 to 9. Only 1 out of 66 coefficients across all balance tests are statistically significant at the 5% level. Overall, these results suggest that our randomization is successful in creating balance across the control and treatment groups.[19]

### 3.4. Health Insurance Enrollment Pattern

Appendix Table A4 shows the number of people by enrollment status in all rounds of survey. We find that long-run enrollment is driven by people who retained their insurance from a previous round. For example, among 865 (=172+33+470+190) people that enrolled in the long run, only 190 (22%) of them enrolled neither in the baseline nor in the short run.

---

[17] In the literature ADL are usually constructed from asking respondents questions about their ability to perform basic daily activities such as self-feeding, ambulation, dressing and undressing etc. The variables used here are derived from the following questions "During the last four weeks did (NAME) have to stop his/her usual activities because of this (illness/injury)" and "For how many days (in the last one month) was name unable to do his/her usual activities".

[18] The main reasons for attrition in the first follow-up survey are deceased (17%), traveled (61%), relocated to other districts (16%), and others (6%). Information on reasons for attrition is not available in the second follow-up survey.

[19] We report additional equality tests and joint orthogonality tests in Appendix Table A3.

Appendix Table A5 presents evidence on general selection pattern in health insurance enrollment by health status and health care utilization regardless of subsidy. Panel A examines selective retention. We restrict the sample to those who enrolled in the baseline (Panels A1 and A2) and in the first follow-up (Panel A3), and examine whether enrollment in the future differs by health status and health care utilization. Similarly, we study selective new enrollment with restricted sample of those that did not enroll in the baseline (Panels B1 and B2), and in the first follow-up (Panel B3). We find some evidence of selective retention. Those who are more ill are more likely to retain their insurance membership. We also find some evidence of selective new enrollment based on health care utilization (Panel B3).

## 4. Estimation Framework

To measure the effects of our intervention on various outcomes, we estimate the following reduced-form intent-to-treat (ITT) effect of each level of subsidy:

$$y_{ihc} = \gamma_0 + \gamma_1 1/3Subsidy_{ihc} + \gamma_2 2/3Subsidy_{ihc} + \gamma_3 FullSubsidy_{ihc} + \theta X_{ihc} + \delta Z_{hc} + \omega V_c + \epsilon_{ihc} \quad (1)$$

where $y_{ihc}$ denotes the outcomes for individual *i* of household *h* in community *c*. The outcomes of interest include NHIS enrollment, health care service utilization, health status, and health behaviors. $X_{ihc}$ denotes a vector of baseline individual covariates, such as indicator variables for age, gender, religion, ethnicity, schooling, and past enrollment to health insurance. Household covariates $Z_{hc}$ include household size and a wealth index indicator (poor third, middle third, and rich third).[20] Community covariates $V_c$ include distance to the nearest health facility and to the NHIS registration center.[21] We also control for a baseline measure of the dependent variable to improve precision. The results are robust when we exclude the baseline controls (results not shown). Estimations employ a linear probability model. For each outcome, we present its short- and long-run estimations.

[20] The wealth index is obtained through a principal components analysis with dwelling characteristics (e.g., number of rooms and bedrooms in the house), enterprise (e.g., ownership of any private non-farm enterprise), livestock (e.g., number of chickens and pigs), and other assets (e.g., motorcycles and bicycles).
[21] In addition, we controlled for indicators for *Subsidy + Campaign, Subsidy + Convenience,* and *Subsidy + Campaign + Convenience.*

We cluster standard errors at the community level to account for possible correlation in the error terms within the same community. [22] We also perform 1,000 draws of a wild-cluster bootstrap percentile *t*-procedure suggested by Cameron et al (2008) to address concerns about small number of clusters, which could lead to downward-biased standard errors (Bertrand et al., 2004; Cameron et al., 2008).

Because we estimate Equation (1) for many different outcome variables in health care utilization and health status domain, a multiple hypothesis testing problem may occur. The probability we incorrectly reject at least one null hypothesis is larger than the conventional significance level. We address this concern using two methods.

First, we group outcome variables into a domain and take the average standardized treatment effect in each domain, as suggested by Kling, Liebman, and Katz (2007) and Finkelstein et al. (2012). For the health care utilization domain, we group five outcome measures including intensive and extensive measures of health facility visits in the last four weeks and last six months and OOP expense incidence. For the health status domain, we group four outcomes including self-rated health status, number of days of illness, inability to perform normal activities, and the number of days lost to illness. Components of the standardized outcome moves in the same direction.

Second, we apply the free step-down resampling procedure to adjust the family-wise error rate, that is, the probability of incorrectly rejecting one or more null hypotheses within a family of hypotheses (Westfall and Young, 1993). Family-wise adjusted *p*-values of each family are obtained from 10,000 simulations of estimations.[23]

[22] To account for correlation within household, we also cluster standard errors at the household level. The results do not change our main conclusion (results available upon request).

[23] These two methods serve different objectives. The first method is relevant for drawing general conclusions about the treatment effects on health care utilization and health status. The second method is more appropriate for examining the treatment effect of a specific outcome belonging to a set of tests.

## 5. Results

### 5.1. Impacts on Insurance Enrollment

Figure 2 shows the enrollment rates of the control and treatment groups at the baseline, short-run follow-up, and long-run follow-up surveys by level of subsidy. In general, it shows that enrollment rate increases with subsidy level in the short run, but the impacts attenuate over time. We observe the largest incremental increase in enrollment rate between receiving zero (control group) and one-third subsidy in the short run, but smaller incremental increases in the subsequent levels of subsidy. In the long run, the treatment group is still more likely to enroll in health insurance, but the differences among the one-third, two-thirds, and full subsidy groups become insignificant.

Table 2 presents the formal regression results. We present robust standard errors in parentheses as well as two-tailed wild cluster bootstrap *p*-values in square brackets. Our results show that the effects on enrollment attenuate but are sustained over time. Column 1 of Panel A shows that overall subsidy intervention increases short-run insurance enrollment by 44.6 percentage points (164%). Long-run enrollment also increases by on average 16.8 percentage points (72%). In terms of the level of subsidy, receiving a one-third, two-thirds, and full subsidy is associated with, respectively, a 36.9, 47.6, and 51.7 percentage points higher likelihood of enrolling in insurance than that of the control group in the short run. Enrollment rates decrease in the long run, but they are still higher than that of the control group. Receiving a one-third, two-thirds, and full subsidy is associated with, respectively, 17.9, 14.3, and 20.4 percentage point higher likelihood of enrolling in insurance than that of the control group. There is no statistical difference in the effects of receiving different subsidy level.

The short-run arc elasticities are large. Overall, when price decreases from \$8.13 to \$0, demand for health insurance increases from 27.2% to 81.0% (arc elasticity is -0.54).[24] This is within range of the elasticity of health insurance in developing countries, such as India, -0.33

[24] We calculate the arc elasticity estimates using the following formula: $[(Y_a - Y_b)/(Y_a + Y_b)]/[(P_a - P_b)/(P_a + P_b)]$, where $Y$ and $P$ denote enrollment rate and price, respectively. The short-run arc elasticity estimates when price increases from \$0 to \$2.67, \$2.67 to \$5.40, and \$5.40 to \$8.13 are 0.04, 0.19, and 2.10, respectively. Comparing the arc elasticity in a zero-price setting to those in other settings could be problematic because the denominator, $(P_a - P_b)/(P_a + P_b)$, is always 1 if $P_b = 0$. Moreover, people tend to treat a zero price not only as a decrease in cost but also as an extra benefit (Shampanier, Mazar, and Ariely, 2007). These results must be interpreted with this caveat.

(Kinnan et al, 2020), and Pakistan, -0.6 (Fischer et al, 2018), and within range of those in developed countries (Pendzialek et al, 2016).[25]

Our finding that a larger subsidy may lead to higher health insurance enrollment corresponds to studies in both developed and developing countries (Finkelstein et al., 2019; Banerjee et al., 2020). However, our finding is contradictory to the special zero price argument suggesting that individuals act as if pricing a good as free not only decreases its cost but also adds to its benefit (Shampanier et al., 2007). For example, several studies find a larger decrease between zero and small non-zero prices in demand for bed nets (Dupas, 2014) and HIV testing (Thornton, 2008).

In contrast, we find a large incremental increase in enrollment between zero and the one-third subsidy (full and two-thirds price) but no difference between the two-thirds and full subsidy (one-third and zero price). One possible explanation for this finding is the framing of the price of health insurance. Unlike Thornton (2008) and Dupas (2014), our subsidy intervention focuses on the level of subsidy instead of the level of price, and, therefore, the largest response to the intervention is found between zero and a small (one-third) subsidy.

### 5.2. Selection into Insurance by Subsidy (Level)

If subsidy attracts people with greater *ex-ante* health risks—more ill and use more health care services—to health insurance, it would threaten financial sustainability of insurance program. We examine selective enrollment into insurance based on complier characteristics analysis framework following Almond and Doyle (2011) and Kim and Lee (2017). In this approach, we compare baseline health and health care utilization characteristics of compliers, always-takers, and never-takers. The impacts we estimate are driven by compliers who enroll in health insurance due to our subsidy intervention. Therefore, we are particularly interested in comparing compliers and never-takers.

[25] The estimated arc elasticity is also close to the elasticity of preventive health products in developing countries, such as -0.6 for chlorine, a disinfectant that prevents water-borne diseases in Zambia (Ashraf, Berry, and Shapiro, 2010), and -0.37 for insecticide-treated bed nets for malaria prevention in Kenya (Cohen and Dupas, 2010). The estimated arc elasticity is also similar to that of preventive health products in developed countries, such as -0.17 and -0.43 for preventive health care in the United States (Newhouse and the Insurance Experiment Group, 1993) and -0.47 for cancer screening in Korea (Kim and Lee, 2017).

To begin, we first define a binary variable $T$, an indicator for whether an individual is assigned to the treatment group (*Subsidy*). Next, we define a binary variable $H$, an indicator for whether an individual is enrolled in health insurance. Lastly, we define $H_T$ as the value $H$ would have if $T$ were either 0 or 1. Hence, $E(X|H_1 = 1)$ presents the mean value characteristics of treated individuals who enrolled in health insurance. Under the assumption of existence of the first stage, monotonicity, and independence, $E(X|H_1 = 1)$ can be written as:

$$E(X|H_1 = 1) = E(X|H_1 = 1, H_0 = 1) \cdot P(H_0 = 1|H_1 = 1) + E(X|H_1 = 1, H_0 = 0) \cdot P(H_0 = 0|H_1 = 1) \quad (2)$$

Equation (2) implies that $E(X|H_1 = 1)$ is a sum of always-takers and compliers components. $E(X|H_1 = 1, H_0 = 0)$ represents the characteristics of compliers. $E(X|H_1 = 1, H_0 = 1) = E(X|H_0 = 1)$ holds from the monotonicity assumption. $P(H_0 = 1)$, the proportion of always-takers, and $P(H_1 = 0)$, the proportion of never-takers, can be directly measured from the sample. $P(H_0 = 1)$, the proportion of always-takers can be thus measured by $P_a$, the proportion of insurance takers in the control group. Similarly, the proportion of never-takers, $P(H_1 = 0)$, can also be measured by $P_b$, the proportion of insurance non-takers in the treatment group. The proportion of compliers is 1- $P_a$ - $P_b$. Therefore, $P(H_0 = 1|H_1 = 1)$ and $P(H_0 = 0|H_1 = 1)$ are $\frac{P_a}{P_c+P_a}$ and $\frac{P_c}{P_c+P_a}$, respectively.[26] By rearranging equation (2), the characteristic of compliers can be calculated as follows:

$$E(X|H_1 = 1, H_0 = 0) = \frac{P_c+P_a}{P_c} \times [E(X|H_1 = 1) - \frac{P_a}{P_c+P_a} \times E(X|H_0 = 1)] \quad (3)$$

Table 3 presents the summary statistics of baseline standardized health status and health care utilization of compliers, always-takers, and the never-takers for short-run selection (Columns 1 to 3) and long-run selection (Columns 7 to 9).[27] We calculate compliers characteristics of any subsidy group in Panel A. We calculate compliers characteristics of partial subsidy group by excluding the full subsidy group from the sample (Panel B). Similarly, we calculate compliers characteristics of full subsidy group by excluding the partial subsidy group from the sample (Panel C). Columns 4 to 6 report the *t*-statistics for the mean comparison of each group in the short run. Columns 10 to 12 report similar statistics in the long run.

[26] The estimated share of compliers, always-takers, and never-takers for different subgroups comparison are as follow: for control vs any subsidy the numbers are 47.4%, 27.1%, and 25.5% in the short run and 24.3%, 23.0%, and 52.7% in the long run; for control vs partial subsidy 44.2%, 27.1%, and 28.6% in the short run and 20.9%, 23%, and 56% in the long run; for control vs full subsidy 50.2%, 27.1%, and 22.6% in the short run and 27%, 23%, and 49.9% in the long run.

[27] We present the results on the components of standardized variables and other variables in Appendix Tables A6 (control vs any subsidy), A7 (control vs partial subsidy), and A8 (control vs full subsidy).

By comparing compliers and never-takers, we find that subsidy intervention attracted people who are more ill both in the short and long run (Columns 5 and 11), and the difference becomes larger in the long run. Similarly, we observe selection based on baseline health status for partial and full subsidy groups in Panels B and C. However, selection based on baseline health care service utilization is only detected among the partial subsidy group.

Next, we explore possible differences in selection between compliers in the partial and full subsidy groups. To do so, we restrict the sample to those who enrolled in health insurance among the full and partial subsidy treatment groups following Kim and Lee (2017). Since we restrict our sample to enrollees in the treatment group, which consists of compliers and always-takers, any difference between full and partial subsidy groups in the restricted sample is due to the compositional changes of compliers.[28]

Table 4 presents the results where we regress standardized measures of baseline health status and health care utilization on an indicator of full subsidy.[29] In each panel, we present results on selection based on health status and health care utilization at baseline. Panels A and B provide results on selection by subsidy level in the short and long run, respectively. We find no evidence on selection by subsidy level based on baseline characteristics.

### 5.3. *Ex-post* Behavioral Responses

In this sub-section, we study impacts on health care service utilization and the role of *ex-post* behavioral responses by investigating correlation between insurance premium and health care utilization (Einav and Finklestein, 2011; Chiappori and Salanie, 2013).

Table 5 presents the effects on utilization of health care services in the short run. Column 6 presents average standardized effects. We report bootstrap and family-wise *p*-values in square and curly brackets, respectively. The long-run effects are presented in Table 6. We find that receiving subsidy leads to an increase in utilization of health care services in the long run (Panel A1). Interestingly, we find increased health care utilization for the partial subsidy group in the long run (Panel A2) even though we do not find significant differential selection and enrollment increases between partial and full subsidy groups. As we do not find difference in selection between compliers in the partial and full subsidy group, the change of health care utilization in the long run might be due to behavioral responses. This behavioral response corresponds to learning-through-experience channel which we discuss in Section 1 because sunk cost fallacy predicts that health care utilization should increase only in the short run when the subsidy still applies—which we do not find.

[28] We impose a reasonable assumption that always-takers in the full and partial subsidy groups are the same.
[29] We present the results on the components of standardized variables in Appendix Tables A9 and A10.

We further investigate learning-through-experience to provide empirical support to our argument. We first restrict the sample to those who enrolled in health insurance among the full and partial subsidy treatment groups. By doing so, we could compare compliers of the partial and full subsidy group as we do in Table 4. We then regress health status and health care utilization measured at the follow-up surveys on an indicator of full subsidy (Table 7). We find that those attracted by partial subsidy are more likely to use health care services in the long run—but not in the short run—than those attracted by full subsidy. These results are consistent with our finding observed in Tables 5 and 6 that health care utilization only increased in the long run and only in the partial subsidy group.

Next, we find limited evidence that health insurance prevents OOP expenses either in the short or long run (Column 5).[30] There are a few possible explanations for this finding. First, as we described earlier, most services are free under the NHIS, but health care providers often charge unauthorized fees as copayments. Second, medicine is often in short supply at the public health centers, and those who receive a diagnosis may purchase medicine from a private pharmacy. Third, those without health insurance often use traditional or herbal medicine which is inexpensive, and therefore, substitution from traditional medicine to formal health care does not decrease OOP expenses.

[30] Again, the size effects in the short- and long-run are not directly comparable because the short- and long-run OOP expenses are constructed differently. In the short run, respondents were asked about more general OOP expenses, but in the long run, OOP expenses only included those related to the treatment of several important illnesses (e.g., malaria, skin diseases, and acute respiratory infection). Specifically, for the short-run OOP expense, we use the individual's response to the following question: "On (NAME's) most recent visit to a health facility, did he/she pay any money from his/her own pocket at a health facility in the last six months?" On the other hand, to construct the long-run OOP expense, we use information on whether individuals made positive OOP expenses in each illness episode (i.e., malaria, acute respiratory infection, and skin diseases) that occurred in the last six months.

**5.4 Impacts on Subjective Health Status.**

Appendix Tables A11 and A12 present the effects on health status in the short and long run (Columns 1 to 5). Column 5 presents average standardized effects. Panel A1 of Table A11 shows that insurance coverage improves health status in the short run. However, Panel A1 of Table A12 shows that the short-run positive health effect seems to disappear in the long run despite increased health insurance enrollment and health care service utilization, as shown in Tables 2 and 6. We even find negative health effects on the number of sick days and daily activities in the long run (Columns 2 to 4 of Table A12).[31]

## 6. Conclusion

Many governments in developing countries aim to increase health insurance enrollment and ultimately achieve universal coverage. To achieve this goal, they offer low sign-up costs and generous benefits, but the enrollment and retention rates remain low. Due to administrative constraints, governments find it difficult to impose mandates and enforce people to enroll in health insurance.

In this paper, we study the long-run impacts of providing an one-time health insurance subsidy of varying levels. We find that the subsidy significantly promoted enrollment in the short run, and while the impacts attenuate, the positive impacts remain three years after the initial intervention implementation. We also find that the subsidy generated selective enrollment. The one-time subsidy attracted people who are *ex-ante* more ill and use more health care services to enroll, though this selection does not vary by subsidy level (partial vs. full). Further, we observe an *ex-post* behavioral response, where health care utilization increases in the long run for the partial subsidy group, not for the full subsidy group. We find empirical evidence that this behavioral response can be explained by stronger learning-through-experience in the partial subsidy group as compared to the full subsidy group.

---

[31] There are two possible mechanism on the lack of long-run health outcomes: moral hazard and perception change on subjective health status. We first investigate individuals' health behaviors – 12 years old or older – regarding the use of bed nets and safe water technologies (Appendix Table A13). We find some suggestive evidence on the decrease in the overall health investments in the full subsidy group, which is not consistent with the results in health utilization and status. This rules out moral hazard as a plausible channel. Second, this could happen when people make frequent contacts with health facilities, learn about the specific symptoms of illnesses, and changed their perception on subjective health status (Dow et al, 1997; Finkelstein, et al, 2012). Also, those who receive a diagnosis could be more aware of the times or periods they were sick. As a result, they are more likely to report being ill. Unfortunately, we are unable to test these conjectures in our data. More research is needed to verify more precise mechanisms through which health insurance enrollment and health care utilization may result in a decline in self-reported health status.

Critics of the Ghanaian NHIS have argued that the scheme is overly generous and financially unsustainable because of the huge percentage of NHIS members under premium exemption without co-payment (Alhassan et al., 2016). Our results suggest that subsidies, especially partial subsidies, could negatively affect the financial sustainability of the social health insurance program. Policy makers should be cautious of the presence of selection and behavioral responses since they are often difficult to predict and, importantly, may endanger the financial stability of an insurance program, especially when mandates are not enforceable.

Taken together, these findings highlight that even though short-run interventions successfully increase health insurance enrollment, their long-run success could depend on selection and behavioral responses. Our findings suggest that as health insurance continues to be introduced in developing counties, both careful enforcement of mandatory health insurance enrollment to prevent selection and establishment of policies to encourage desirable health behaviors need to be considered.

**Figures and Tables**

Figure 1: Study Design

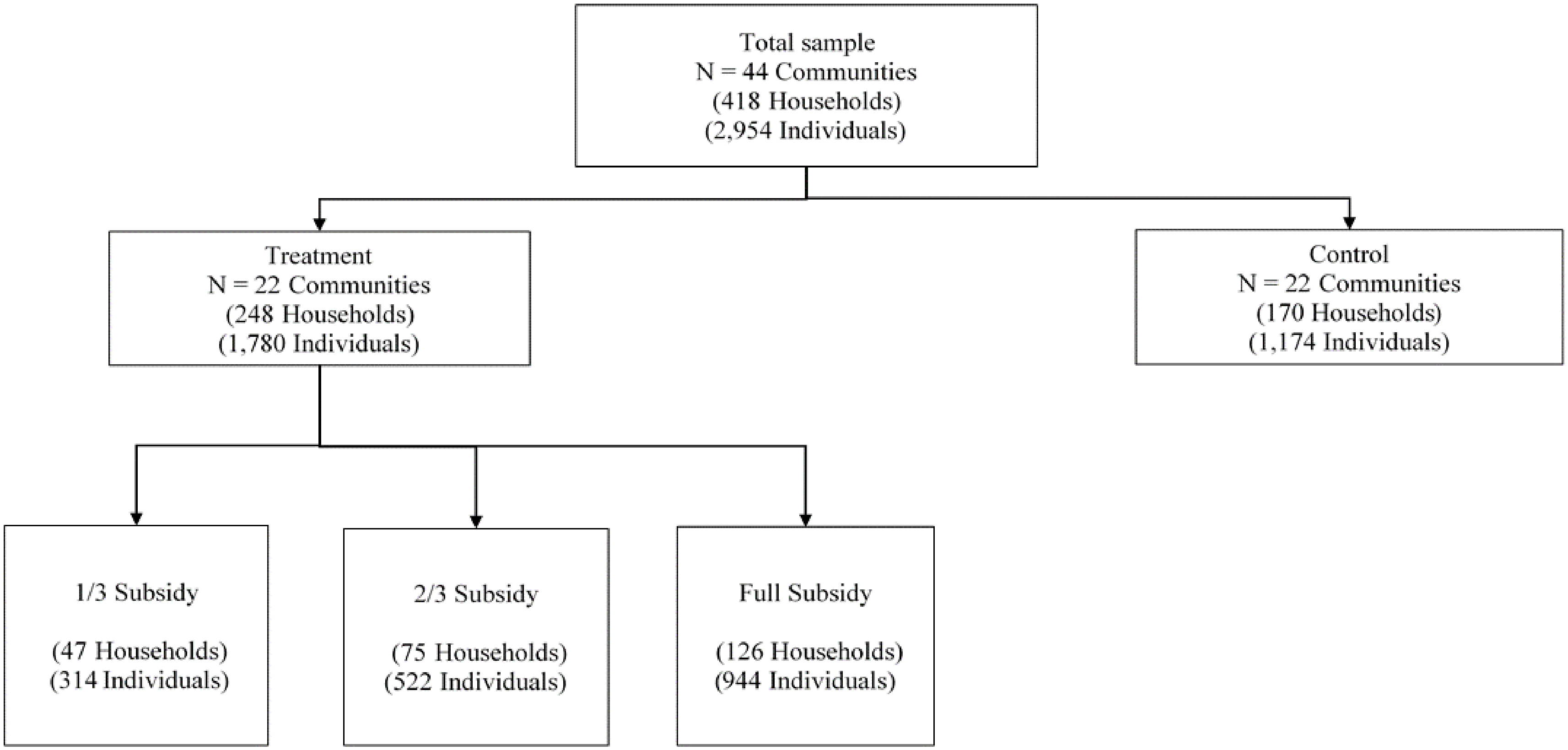

Figure 2: Enrollment Rate by Subsidy Level at Baseline, Short Run, and Long Run

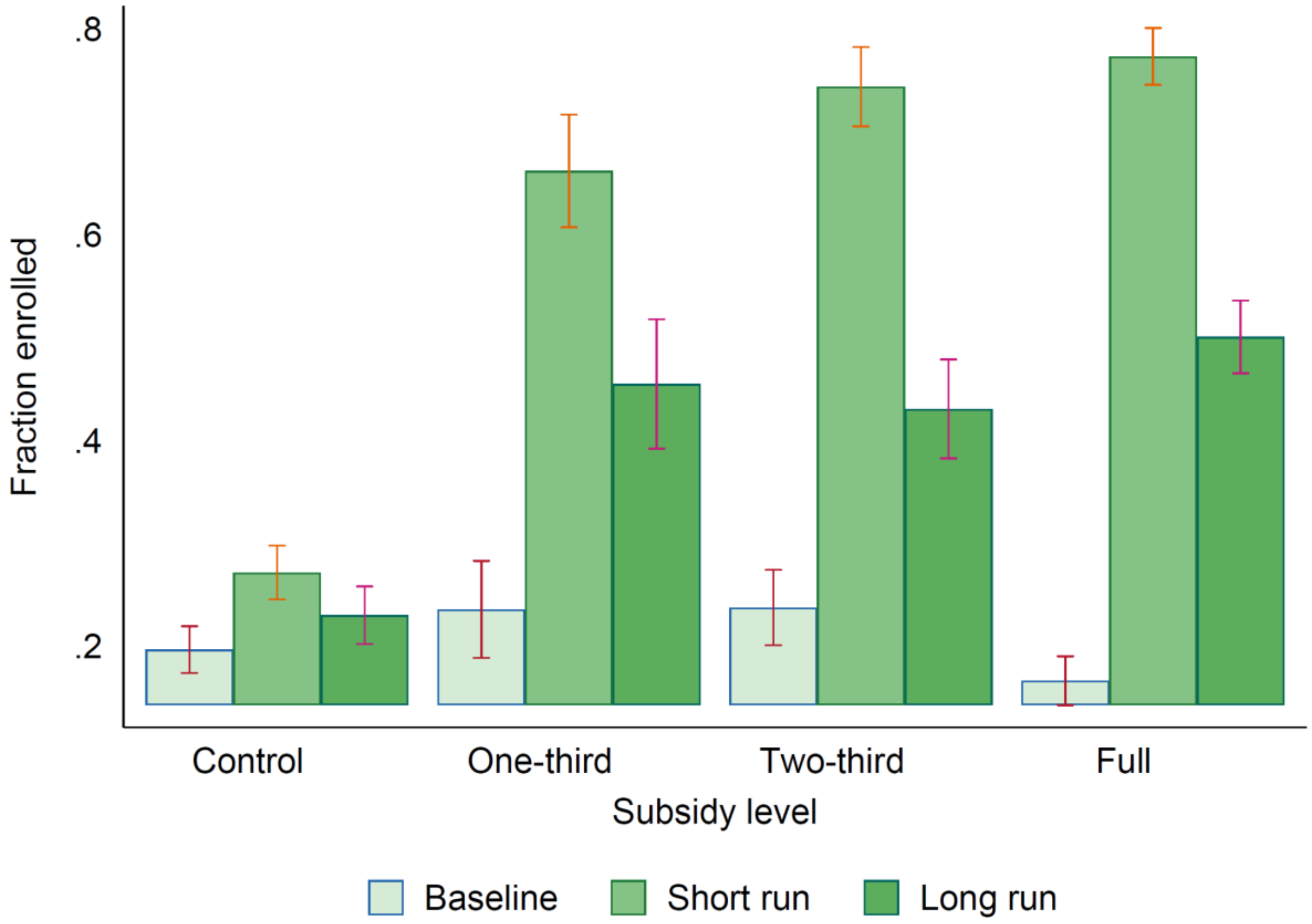

Notes: This figure shows means of enrollment rates of each subsidy-level group at baseline, short run, and long run. Sample includes those who received subsidy and the control group. The vertical lines indicate 95% confidence intervals. Standard errors are not clustered.

Table 1: Baseline Summary Statistics and Randomization Balance

| | Mean | | Difference between subsidy level and control | | | | Difference between each subsidy level | | |
|---|---|---|---|---|---|---|---|---|---|
| | All | Control | One-third | Two-thirds | Full | N | One-Third vs Two- | One-Third vs Full | Two-Thirds vs Full |
| | (1) | (2) | (3) | (4) | (5) | (6) | (7) | (8) | (9) |
| ***Panel A: Control variables*** | | | | | | | | | |
| Age (years) | 23.780 | 24.313 | 1.180 | -0.775 | -1.620 | 2,954 | 1.955 | 2.800 | 0.844 |
| Male | 0.481 | 0.475 | 0.009 | -0.010 | 0.022 | 2,954 | 0.019 | -0.013 | -0.031 |
| Christian | 0.417 | 0.373 | 0.073 | 0.102 | 0.058 | 2,954 | -0.029 | 0.015 | 0.044 |
| Dagaaba ethnic group | 0.517 | 0.458 | 0.153 | 0.208 | 0.017 | 2,954 | -0.055 | 0.136 | 0.191 |
| Has some formal education | 0.335 | 0.337 | -0.022 | -0.015 | 0.009 | 2,954 | -0.007 | -0.031 | -0.025 |
| Household size | 8.703 | 8.454 | -0.187 | 0.051 | 0.813 | 2,953 | -0.238 | -0.999 | -0.761 |
| Household assets (principal component score) | 0.601 | 0.266 | 0.580 | 0.269 | 0.705** | 2,953 | 0.311 | -0.126 | -0.436 |
| Distance to district insurance office (km) | 20.010 | 20.37 | 4.347 | 3.447 | -4.466 | 2,954 | 0.900 | 8.812 | 7.912 |
| Distance to nearest health facility (km) | 5.394 | 5.166 | 0.221 | -0.687 | 1.017 | 2,954 | 0.908 | -0.796 | -1.704** |
| Ever enrolled in NHIS | 0.358 | 0.302 | 0.179** | 0.084 | 0.071 | 2,954 | 0.096 | 0.108 | 0.012 |
| ***Panel B: Baseline values of outcome variables*** | | | | | | | | | |
| Enrollment | 0.198 | 0.197 | 0.039 | 0.041 | -0.030 | 2,954 | -0.002 | 0.069 | 0.071 |
| Ill in the last four weeks | 0.123 | 0.105 | 0.039 | 0.048 | 0.016 | 2,954 | -0.010 | 0.023 | 0.032 |
| Number of sick days in the last four weeks | 0.918 | 0.846 | 0.505 | 0.208 | -0.056 | 2,927 | 0.296 | 0.560 | 0.264 |
| Could not do normal activities in the last four weeks | 0.076 | 0.060 | 0.011 | 0.039 | 0.023 | 2,919 | -0.028 | -0.012 | 0.016 |
| No. of days could not perform normal activities in the last four weeks | 0.544 | 0.480 | 0.134 | 0.138 | 0.079 | 2,815 | -0.004 | 0.055 | 0.060 |
| Visited health facility in the last four weeks | 0.039 | 0.036 | 0.033 | 0.023 | -0.015 | 2,435 | 0.010 | 0.048 | 0.038 |
| Visited health facility in the last six months | 0.074 | 0.074 | 0.025 | 0.008 | -0.014 | 2,954 | 0.016 | 0.038 | 0.022 |
| Number of health facility visit in the last four weeks | 0.066 | 0.063 | 0.062 | 0.042 | -0.036 | 2,443 | 0.020 | 0.098 | 0.078* |
| Visited health facility for malaria treatment in the last four weeks | 0.010 | 0.011 | -0.004 | 0.002 | -0.004 | 2,435 | -0.006 | -0.0005 | 0.006 |
| Made out of pocket expense in the last six months | 0.136 | 0.133 | -0.009 | 0.059 | -0.021 | 2,954 | -0.067 | 0.012 | 0.079* |
| Standardized health care utilization | 0.000 | -0.001 | 0.014 | 0.018 | -0.013 | 13,221 | -0.004 | 0.027 | 0.017 |
| Standardized health status | 0.000 | -0.011 | 0.026 | 0.029 | 0.011 | 8,661 | -0.003 | 0.014 | 0.031* |

Notes: Columns 1 and 2 report mean of all respondents and control group. Columns 3 to 5 present results from regressions of each variable on control and subsidy level indicators (1/3, 2/3, and full). Column 6 reports total number of observations. Columns 7 to 9 present results from separate regressions of each variable on one-third and two-thirds subsidy levels (Column 7), on one-third and full subsidy levels (Column 8), and on two-thirds and full subsidy levels (Column 9). Robust standard errors are clustered at community level. *, **, and *** denote statistical significance at 10 %, 5 %, and 1 %. levels, respectively.

Table 2: Effects of Subsidy (Level) on NHIS Enrollment

| | Enrollment | |
|---|---|---|
| | Short-run | Long-run |
| | (1) | (2) |
| **Panel A** | | |
| Any Subsidy | 0.446*** | 0.168* |
| | (0.054) | (0.084) |
| | [0.000] | [0.103] |
| R-squared | 0.359 | 0.186 |
| **Panel B** | | |
| Partial subsidy | 0.430*** | 0.158* |
| | (0.055) | (0.082) |
| | [0.000] | [0.117] |
| Full subsidy | 0.508*** | 0.207** |
| | (0.060) | (0.102) |
| | [0.000] | [0.125] |
| R-squared | 0.361 | 0.187 |
| **Panel C** | | |
| 1/3 subsidy | 0.369*** | 0.179** |
| | (0.073) | (0.087) |
| | [0.001] | [0.115] |
| 2/3 subsidy | 0.476*** | 0.143 |
| | (0.060) | (0.087) |
| | [0.000] | [0.156] |
| Full subsidy | 0.517*** | 0.204* |
| | (0.057) | (0.102) |
| | [0.000] | [0.105] |
| R-squared | 0.364 | 0.188 |
| Number of observations | 2,785 | 2,257 |
| Mean | 0.555 | 0.383 |
| Control group mean | 0.272 | 0.233 |
| **P-values on test of equality:** | | |
| Partial subsidy = Full subsidy | 0.120 | 0.435 |
| 1/3 subsidy = 2/3 subsidy | 0.122 | 0.585 |
| 1/3 subsidy = Full subsidy | 0.014 | 0.702 |
| 2/3 subsidy = Full subsidy | 0.448 | 0.408 |

Notes: All regressions include a standard set of covariates (individual, household, and community) and baseline measure of dependent variable. *P*-values for the equality of effect estimates for various pairs of treatment groups are also presented. Robust standard errors clustered at community level are reported in parentheses. Wild-cluster bootstrap-t *p*-values are reported in square brackets. *, **, and *** denote statistical significance at 10 %, 5 %, and 1 % levels, respectively.

Table 3: Selection by Subsidy (Level): Characteristics of Compliers, Always Takers, and Never Takers

| | Short run | | | | | | Long run | | | | | |
|---|---|---|---|---|---|---|---|---|---|---|---|---|
| | Mean | | | t-stat | | | Mean | | | t-stat | | |
| | Complier | Always | Never | C=A | C=N | A=N | Complier | Always | Never | C=A | C=N | A=N |
| | (1) | (2) | (3) | (4) | (5) | (6) | (7) | (8) | (9) | (10) | (11) | (12) |
| **Panel A: Control vs Any subsidy** | | | | | | | | | | | | |
| Proportion | 47.40 | 27.10 | 25.50 | | | | 24.26 | 23.01 | 52.73 | | | |
| Baseline standardized health status | 0.0100 | -0.0073 | -0.0032 | 5.19 | 4.91 | -0.95 | 0.0147 | 0.0141 | -0.0063 | 0.09 | 9.61 | 2.91 |
| Baseline standardized health care utilization | 0.0011 | 0.0027 | -0.0085 | -0.28 | 2.47 | 1.59 | -0.0050 | 0.0277 | -0.0099 | -3.77 | 1.66 | 4.10 |
| **Panel B: Control vs Partial subsidy** | | | | | | | | | | | | |
| Proportion | 44.22 | 27.15 | 28.63 | | | | 20.96 | 23.01 | 56.03 | | | |
| Baseline standardized health status | 0.0142 | -0.0073 | 0.0024 | 6.45 | 2.88 | -1.85 | 0.0229 | 0.0141 | -0.0030 | 1.31 | 7.43 | 2.28 |
| Baseline standardized health care utilization | 0.0250 | 0.0027 | 0.0042 | 3.77 | 3.37 | -0.18 | 0.0240 | 0.0277 | 0.0024 | -0.42 | 4.28 | 2.52 |
| **Panel C: Control vs Full subsidy** | | | | | | | | | | | | |
| Proportion | 50.20 | 27.15 | 22.65 | | | | 27.05 | 23.01 | 49.94 | | | |
| Baseline standardized health status | 0.0068 | -0.0073 | -0.0096 | 4.22 | 4.72 | 0.46 | 0.0094 | 0.0141 | -0.0094 | -0.71 | 7.01 | 3.27 |
| Baseline standardized health care utilization | -0.0177 | 0.0027 | -0.0241 | -3.46 | 1.54 | 3.72 | -0.0239 | 0.0277 | -0.0221 | -5.96 | -0.61 | 5.42 |

Note: This table presents the mean baseline characteristics of compliers, always takers, and never takers, which are estimated from Equation (3). Columns 4-6 and 10-12 present the *t*-statistics from the two-sample *t*-test comparing compliers with always takers, compliers with never takers, and always takers with never takers, respectively.

Table 4: Selection by Subsidy Level (Partial vs Full Subsidy): Baseline Standardized Health Status and Health Care Utilization among Enrollees

| Independent variable: Received full subsidy | Coefficient (1) | Std. Error (2) | N (3) | R-squared (4) |
|---|---|---|---|---|
| **Sample** | Enrolled in the short run | | | |
| **Panel A** | | | | |
| Baseline standardized health status | 0.011 | (0.020) | 7,314 | 0.006 |
| Baseline standardized health care utilization | -0.019 | (0.022) | 5,759 | 0.007 |
| **Sample** | Enrolled in the long run | | | |
| **Panel B** | | | | |
| Baseline standardized health status | 0.019 | (0.034) | 3,944 | 0.013 |
| Baseline standardized health care utilization | 0.011 | (0.032) | 3,092 | 0.023 |

Notes: This table reports estimation results of running separate regressions of standardized health status and health care utilization on an indicator variable that takes value of one if receiving full subsidy and zero if receiving partial subsidy. We control for indicators of other interventions involving subsidy: *Subsidy + Campaign, Subsidy + Convenience,* and *Subsidy + Campaign + Convenience.* Sample is restricted to enrollees who received partial and full subsidy. Panels A and B summarize regression results when sample is restricted to those who enrolled in the short run. Panel C summarizes results when sample is restricted to those who enrolled in both short and long run, while Panel D is restricted to those who enrolled in all rounds. Robust standard errors clustered at community level reported in parentheses. *, **, and *** denote statistical significance at 10 %, 5 %, and 1 % level respectively.

Table 5: Effects on Healthcare Services Utilization (Short Run)

| | Short run | | | | | |
|---|---|---|---|---|---|---|
| | Visited health facility in last four weeks | Visited health facility in last six months | # of visits in last four weekss | Visited facility for malaria treatment in the last four weeks | Made out-of-pocket for health service in the last six months | Standardized treatment effects |
| | (1) | (2) | (3) | (4) | (5) | (6) |
| **Panel A: ITT results** | | | | | | |
| **Panel A1** | | | | | | |
| Any subsidy | -0.004 | 0.005 | -0.0002 | 0.006 | -0.009 | 0.001 |
| | (0.012) | (0.020) | (0.023) | (0.008) | (0.016) | (0.010) |
| | [0.725] | [0.802] | [0.992] | [0.457] | [0.606] | |
| | {0.973} | {0.973} | {0.993} | {0.912} | {0.937} | |
| R-squared | 0.106 | 0.132 | 0.066 | 0.074 | 0.094 | 0.058 |
| **Panel A2** | | | | | | |
| Partial subsidy | -0.007 | 0.007 | -0.002 | 0.008 | -0.009 | -0.000 |
| | (0.013) | (0.019) | (0.024) | (0.009) | (0.015) | (0.010) |
| | [0.606] | [0.731] | [0.929] | [0.400] | [0.620] | |
| | {0.947} | {0.947} | {0.947} | {0.873} | {0.947} | |
| Full subsidy | 0.009 | -0.001 | 0.008 | 0.000 | -0.012 | 0.004 |
| | (0.022) | (0.031) | (0.038) | (0.012) | (0.023) | (0.016) |
| | [0.700] | [0.965] | [0.842] | [0.981] | [0.606] | |
| | {0.984} | {0.999} | {0.997} | {0.999} | {0.981} | |
| R-squared | 0.107 | 0.132 | 0.066 | 0.074 | 0.094 | 0.058 |
| **Panel A3** | | | | | | |
| 1/3 subsidy | 0.001 | 0.004 | -0.010 | 0.009 | -0.014 | 0.001 |
| | (0.017) | (0.022) | (0.024) | (0.011) | (0.014) | (0.011) |
| | [0.958] | [0.835] | [0.698] | [0.480] | [0.362] | |
| | {0.978} | {0.978} | {0.956} | {0.885} | {0.832} | |
| 2/3 subsidy | -0.014 | 0.008 | 0.004 | 0.007 | -0.004 | -0.001 |
| | (0.014) | (0.023) | (0.029) | (0.010) | (0.020) | (0.011) |
| | [0.381] | [0.738] | [0.877] | [0.485] | [0.845] | |
| | {0.848} | {0.975} | {0.976} | {0.900} | {0.976} | |
| Full subsidy | 0.007 | -0.001 | 0.010 | 0.000 | -0.011 | 0.004 |
| | (0.021) | (0.031) | (0.037) | (0.012) | (0.023) | (0.016) |
| | [0.718] | [0.976] | [0.812] | [0.981] | [0.676] | |
| | {0.994} | {0.999} | {0.994} | {0.999} | {0.985} | |
| R-squared | 0.107 | 0.132 | 0.066 | 0.074 | 0.094 | 0.058 |
| Number of observations | 2,130 | 2,710 | 2,124 | 2,252 | 2,805 | 11,008 |
| Control group mean | 0.038 | 0.102 | 0.032 | 0.019 | 0.046 | -0.011 |
| **P-values on test of equality:** | | | | | | |
| Partial subsidy = Full subsidy | 0.484 | 0.745 | 0.777 | 0.541 | 0.827 | 0.783 |
| 1/3 subsidy = 2/3 subsidy | 0.393 | 0.870 | 0.559 | 0.905 | 0.550 | 0.813 |
| 1/3 subsidy = Full subsidy | 0.762 | 0.847 | 0.571 | 0.483 | 0.838 | 0.856 |
| 2/3 subsidy = Full subsidy | 0.426 | 0.731 | 0.891 | 0.621 | 0.721 | 0.768 |

Notes: Panel A reports ITT results. Panels A1, A2, and A3 report the effects of receiving any subsidy, partial and full subsidy, and each subsidy level (1/3, 2/3, and full) on health care utilization in the short run. All regressions include a standard set of covariates (individual, household, and community) and baseline measure of dependent variable. Standardized treatment effects are reported in Column 6. *P*-values for the equality of effect estimates for various pairs of treatment groups are also presented. Robust standard errors clustered at community level are reported in parentheses. Wild-cluster bootstrap-t *p*-values are reported in square brackets. Family-wise *p*-values are reported in curly brackets. *, **, and *** denote statistical significance at 10 %, 5 %, and 1 % levels, respectively.

Table 6: Effects on Healthcare Services Utilization (Long Run)

| | Long run | | | | | |
|---|---|---|---|---|---|---|
| | Visited health facility in last four weeks | Visited health facility in last six months | # of visits in last four weekss | Visited facility for malaria treatment in the last four weeks | Made out-of-pocket for health service in the last six months | Standardized treatment effects |
| | (1) | (2) | (3) | (4) | (5) | (6) |
| **Panel A: ITT results** | | | | | | |
| **Panel A1** | | | | | | |
| Any subsidy | 0.031** | 0.080*** | 0.022* | 0.020 | 0.002 | 0.030** |
| | (0.014) | (0.026) | (0.013) | (0.015) | (0.007) | (0.013) |
| | [0.054] | [0.010] | [0.141] | [0.264] | [0.767] | |
| | {0.207} | {0.088} | {0.310} | {0.432} | {0.776} | |
| R-squared | 0.078 | 0.088 | 0.065 | 0.067 | 0.090 | 0.064 |
| **Panel A2** | | | | | | |
| Partial subsidy | 0.047*** | 0.106*** | 0.037*** | 0.034** | 0.013 | 0.047*** |
| | (0.013) | (0.024) | (0.012) | (0.014) | (0.009) | (0.012) |
| | [0.001] | [0.000] | [0.003] | [0.028] | [0.239] | |
| | {0.068} | {0.032} | {0.085} | {0.157} | {0.272} | |
| Full subsidy | -0.030 | -0.015 | -0.035 | -0.036* | -0.037* | -0.038* |
| | (0.019) | (0.051) | (0.022) | (0.020) | (0.022) | (0.022) |
| | [0.202] | [0.824] | [0.203] | [0.175] | [0.063] | |
| | {0.636} | {0.822} | {0.636} | {0.609} | {0.636} | |
| R-squared | 0.095 | 0.103 | 0.080 | 0.085 | 0.103 | 0.080 |
| **Panel A3** | | | | | | |
| 1/3 subsidy | 0.018 | 0.082** | 0.014 | 0.018 | 0.026 | 0.034 |
| | (0.015) | (0.033) | (0.017) | (0.016) | (0.027) | (0.021) |
| | [0.329] | [0.051] | [0.511] | [0.372] | [0.687] | |
| | {0.640} | {0.410} | {0.677} | {0.649} | {0.677} | |
| 2/3 subsidy | 0.070*** | 0.124*** | 0.055*** | 0.048*** | 0.004 | 0.057*** |
| | (0.016) | (0.032) | (0.016) | (0.018) | (0.012) | (0.015) |
| | [0.000] | [0.000] | [0.008] | [0.016] | [0.792] | |
| | {0.031} | {0.040} | {0.057} | {0.113} | {0.810} | |
| Full subsidy | -0.027 | -0.012 | -0.032 | -0.034 | -0.039 | -0.037 |
| | (0.019) | (0.052) | (0.022) | (0.020) | (0.024) | (0.023) |
| | [0.277] | [0.884] | [0.228] | [0.222] | [0.102] | |
| | {0.647} | {0.867} | {0.647} | {0.646} | {0.647} | |
| R-squared | 0.100 | 0.105 | 0.084 | 0.088 | 0.105 | 0.081 |
| Number of observations | 2,228 | 2,688 | 2,231 | 2,228 | 2,688 | 11,140 |
| Control group mean | 0.017 | 0.050 | 0.036 | 0.010 | 0.013 | -0.021 |
| **P-values on test of equality:** | | | | | | |
| Partial subsidy = Full subsidy | 0.000 | 0.008 | 0.003 | 0.001 | 0.086 | 0.001 |
| 1/3 subsidy = 2/3 subsidy | 0.012 | 0.319 | 0.092 | 0.155 | 0.533 | 0.384 |
| 1/3 subsidy = Full subsidy | 0.061 | 0.088 | 0.081 | 0.044 | 0.196 | 0.060 |
| 2/3 subsidy = Full subsidy | 0.000 | 0.004 | 0.002 | 0.001 | 0.020 | 0.000 |

Notes: Panel A reports ITT results. Panels A1, A2, and A3 report the effects of receiving any subsidy, partial and full subsidy, and each subsidy level (1/3, 2/3, and full) on health care utilization in the long run. All regressions include a standard set of covariates (individual, household, and community) and baseline measure of dependent variable. Standardized treatment effects are reported in Column 6. *P*-values for the equality of effect estimates for various pairs of treatment groups are also presented. Robust standard errors clustered at community level are reported in parentheses. Wild-cluster bootstrap-t *p*-values are reported in square brackets. Family-wise *p*-values are reported in curly brackets. *, **, and *** denote statistical significance at 10 %, 5 %, and 1 % levels, respectively.

Table 7: Ex-post Behavioral Response (Partial vs Full Subsidy): Effects on Follow-up Standardized Health Care Utilization among Enrollees

| Independent variable: Received full subsidy | Coefficient (1) | Std. Error (2) | N (3) | R-squared (4) |
|---|---|---|---|---|
| **Sample** | Enrolled in the short run | | | |
| **Panel A** | | | | |
| Short-run standardized health care utilization | -0.007 | (0.008) | 5,967 | 0.009 |
| **Sample** | Enrolled in the long run | | | |
| **Panel B** | | | | |
| Short-run standardized health care utilization | -0.012 | (0.014) | 3,211 | 0.007 |
| Long-run standardized health care utilization | -0.116*** | (0.034) | 3,370 | 0.038 |

Notes: This table reports estimation results of running separate regressions of follow-up standardized health care utilization on an indicator variable that takes value of one if receiving full subsidy and zero if receiving partial subsidy. We control for indicators of other interventions involving subsidy: *Subsidy + Campaign*, *Subsidy + Convenience*, and *Subsidy + Campaign + Convenience*. Sample is restricted to enrollees who received partial and full subsidy. Panels A and B summarize regression results when sample is restricted to those who enrolled in the short and long run, respectively. Robust standard errors clustered at community level reported in parentheses. *, **, and *** denote statistical significance at 10 %, 5 %, and 1 % level respectively.

## Appendix A

Figure A.1. Ghana and Wa West District Map

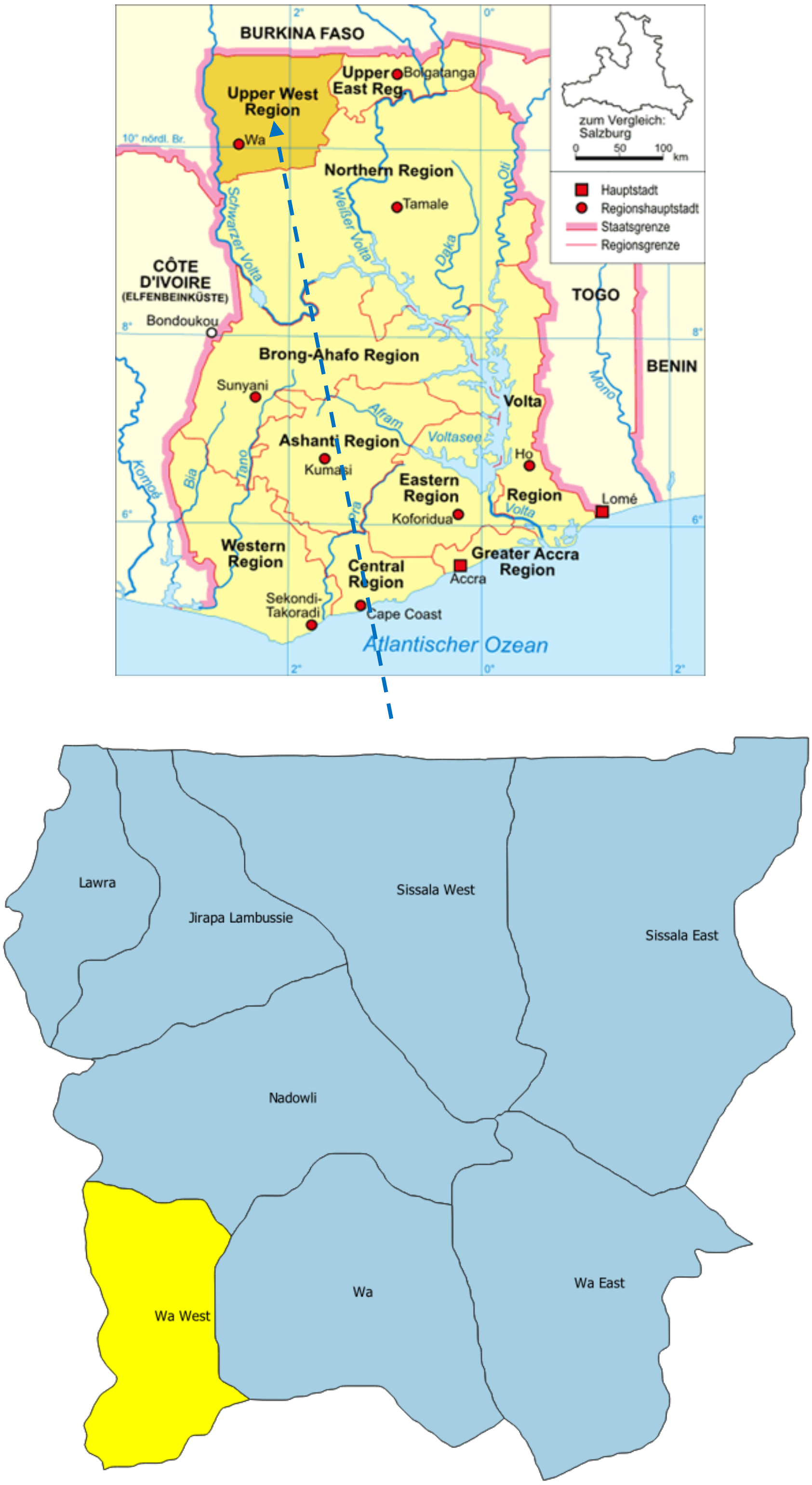

Note: This map shows Ghana (upper panel) and the Upper West region of Ghana (lower panel), which includes Wa West district (highlighted).

Table A1: NHIS Coverage

| Included Services | Exclusion List |
|---|---|
| 1 Out-Patient Services | 1 Rehabilitation other than physiotherapy |
| i) General and specialized consultation and review | |
| ii) Requested investigation (including laboratory investigations, x-rays and ultrasound scanning) | 2 Appliances and protheses including optical aids, hearing aids, othopedic aids and dentures |
| iii) Medication (prescription drugs on the NHIS Drug List) | |
| iv) HIV/AIDS symptomatic treatment for opportunistic infection | |
| v) Out-patient/Day Surgery Operations including hernia repairs, incision and drainage, hemorrhoidectomy | 3 Cosmetic surgeries and aesthetic treatment |
| vi) Out-patient physiotherapy | |
| | 4 HIV retroviral drugs |
| 2 In-Patient Services | |
| i) General and specialist in-patient care | 5 Assisted reproduction eg artificial insemination and gynecological hormone replacement therapy |
| ii) Requested investigations | |
| iii) Medication (prescription drugs on NHIS Drug List) | |
| iv) Cervical and Breast Cancer Treatment | |
| v) Surgical Operations | 6 Echocardiography |
| vi) In-patient physiotherapy | |
| vii) Accommodation in general ward | 7 Photography |
| viii) Feeding (where available) | |
| | 8 Angiography |
| 3 Oral Health Services | |
| i) Pain relief which includes incision and drainage, tooth extraction and temporary relief | 9 Orthotics |
| ii) Dental restoration which includes simple amalgam fillings and temporary dressing | 10 Dialysis for chronic renal failure |
| | 11 Heart and brain surgery other than those resulting from accident |
| 4 Eye Care Services | |
| i) Refraction, visual fields and A-Scan | |
| ii) Keratometry | 12 Cancer treatment other than cervical ad breast cancer |
| iii) Cataract removal | |
| iv) Eye lid surgery | |
| | 13 Organ transplating |
| 5 Maternity Care | |
| i) Antenatal care | 14 All drugs that not listed on the NHIS Drug List |
| ii) Deliveries (normal and assisted) | |
| iii) Caesarian section | |
| iv) Postnatal care | 15 Diagnosis and treatment abroad |
| 6 Emergencies | 16 Medical examinations for purposes of visa applications, Campaign and institutional driving license |
| i) Medical emergencies | |
| ii) Surgical emergencies including brain surgery due to accidents | |
| iii) Pediatric emergencies | 17 VIP ward accommodation |
| iv) Obstetric and gynecological emergencies | |
| v) Road traffic accidents | 18 Mortuary Services |
| vi) Industrial and workplace accidents | |
| vii) Dialysis for acute renal failure | |

Source: NHIA (2011)

Table A2: Attrition

| | Short run | Long run |
|---|---|---|
| | (1) | (2) |
| **Panel A** | | |
| Any subsidy | 0.006 | -0.047 |
| | (0.024) | (0.036) |
| | [0.818] | [0.224] |
| R-squared | 0.133 | 0.102 |
| **Panel B** | | |
| Partial subsidy | 0.004 | -0.042 |
| | (0.025) | (0.038) |
| | [0.895] | [0.302] |
| Full subsidy | 0.013 | -0.065 |
| | (0.044) | (0.053) |
| | [0.809] | [0.311] |
| R-squared | 0.144 | 0.104 |
| **Panel C** | | |
| 1/3 subsidy | -0.005 | -0.039 |
| | (0.043) | (0.051) |
| | [0.924] | [0.533] |
| 2/3 subsidy | 0.011 | -0.045 |
| | (0.024) | (0.042) |
| | [0.664] | [0.314] |
| Full subsidy | 0.014 | -0.066 |
| | (0.044) | (0.051) |
| | [0.779] | [0.298] |
| R-squared | 0.144 | 0.104 |
| Mean | 0.05 | 0.21 |
| Number of observations | 2953 | 2953 |

Notes: Dependent variable is a binary variable indicating whether an individual had been attrited in the short- and long-run follow-up surveys. All regressions include a standard set of covariates (individual, household, and community). Robust standard errors clustered at community level reported in parentheses. *, **, and *** denote statistical significance at 10 %, 5 %, and 1 % levels, respectively.

Table A3: Additional Balance Tests

| | Mean | | | | | N | P-values of difference |
|---|---|---|---|---|---|---|---|
| | All | Control | One-Third | Two-thirds | Full | | One-third = Two-thirds = Full = Control |
| | (1) | (2) | (3) | (4) | (5) | (6) | (7) |
| ***Panel A: Control variables*** | | | | | | | |
| Age (years) | 23.780 | 24.313 | 25.494 | 23.538 | 22.694 | 2,954 | 0.390 |
| Male | 0.481 | 0.475 | 0.484 | 0.466 | 0.497 | 2,954 | 0.578 |
| Christian | 0.417 | 0.373 | 0.446 | 0.475 | 0.431 | 2,954 | 0.801 |
| Dagaaba ethnic group | 0.517 | 0.458 | 0.611 | 0.667 | 0.476 | 2,954 | 0.370 |
| Has some formal education | 0.335 | 0.337 | 0.315 | 0.322 | 0.346 | 2,954 | 0.976 |
| Household size | 8.703 | 8.454 | 8.268 | 8.506 | 9.267 | 2,953 | 0.517 |
| Household assets (principal component score) | 0.601 | 0.266 | 0.846 | 0.535 | 0.971 | 2,953 | 0.061 |
| Distance to district insurance office (km) | 20.010 | 20.37 | 24.717 | 23.817 | 15.904 | 2,954 | 0.303 |
| Distance to nearest health facility (km) | 5.394 | 5.166 | 5.388 | 4.48 | 6.184 | 2,954 | 0.149 |
| Ever enrolled in NHIS | 0.358 | 0.302 | 0.481 | 0.385 | 0.373 | 2,954 | 0.090 |
| ***Panel B: Baseline values of outcome variables*** | | | | | | | |
| Enrollment | 0.198 | 0.197 | 0.236 | 0.238 | 0.166 | 2,954 | 0.690 |
| Ill in the last four weeks | 0.123 | 0.105 | 0.143 | 0.153 | 0.121 | 2,954 | 0.532 |
| Number of sick days in the last four weeks | 0.918 | 0.846 | 1.350 | 1.054 | 0.790 | 2,927 | 0.565 |
| Could not do normal activities in the last four weeks | 0.076 | 0.060 | 0.071 | 0.099 | 0.084 | 2,919 | 0.428 |
| No. of days could not perform normal activities in the last four weeks | 0.544 | 0.480 | 0.614 | 0.619 | 0.559 | 2,815 | 0.867 |
| Visited health facility in the last four weeks | 0.039 | 0.036 | 0.069 | 0.059 | 0.021 | 2,435 | 0.301 |
| Visited health facility in the last six months | 0.074 | 0.074 | 0.099 | 0.082 | 0.060 | 2,954 | 0.639 |
| Number of health facility visit in the last four weeks | 0.066 | 0.063 | 0.125 | 0.105 | 0.027 | 2,443 | 0.117 |
| Visited health facility for malaria treatment in the last four weeks | 0.010 | 0.011 | 0.007 | 0.013 | 0.007 | 2,435 | 0.925 |
| Made out of pocket expense in the last six months | 0.136 | 0.133 | 0.124 | 0.192 | 0.112 | 2,954 | 0.306 |
| **P-values of joint orthogonality test** | | | | | | | |
| Control variables | | | | | | | 0.425 |
| Outcomes | | | | | | | 0.991 |

Notes: Columns 1 to 5 report mean of baseline characteristics of all respondents, control group, and treatment groups. Column 6 reports total number of observations. Column 7 reports the *p*-value from balance tests for each characteristic. In the last two rows we report a joint test of orthogonality for sets of outcomes and covariates separately. To perform the joint test of orthogonality we regress the treatment variable that takes on all covariates and we then test the joint null hypothesis that all covariates have a zero coefficient. We estimate a categorical logit model because the treatment variable has more than one value.

Table A4: Breakdown of Enrollment Rate in Each Round

| Baseline | | | Short run | | | Long run | | |
|---|---|---|---|---|---|---|---|---|
| | N | % | | N | % | | N | % |
| Enrolled | 586 | 19.84 | Enrolled | 394 | 67.24 | Enrolled | 172 | 43.65 |
| | | | | | | Not enrolled | 165 | 41.88 |
| | | | | | | Missing | 57 | 14.47 |
| | | | Not Enrolled | 141 | 24.06 | Enrolled | 33 | 23.40 |
| | | | | | | Not enrolled | 81 | 57.45 |
| | | | Missing | 51 | 8.70 | Missing | 27 | 19.15 |
| Not enrolled | 2,368 | 80.16 | Enrolled | 1,153 | 48.69 | Enrolled | 470 | 40.76 |
| | | | | | | Not enrolled | 482 | 41.80 |
| | | | | | | Missing | 201 | 17.43 |
| | | | Not enrolled | 1,097 | 46.33 | Enrolled | 190 | 17.32 |
| | | | | | | Not enrolled | 664 | 60.53 |
| | | | Missing | 118 | 4.98 | Missing | 243 | 22.15 |
| | | | Enrolled in short run | 1,547 | 55.55 | Enrolled | 642 | 41.50 |
| | | | | | | Not enrolled | 647 | 41.82 |
| | | | | | | Missing | 258 | 16.68 |
| | | | Not enrolled in short run | 1,238 | 44.45 | Enrolled | 223 | 18.01 |
| | | | | | | Not enrolled | 745 | 60.18 |
| | | | | | | Missing | 270 | 21.81 |

Notes: This table reports the number of people and proportion of enrollment rate in each round. The number of missing observations on enrollment status between rounds are either caused by attrition or missing enrollment information from the non-attrited respondents.

Table A5: Selective Retention and New Enrollment by Health Status and Health Care Utilization

| | Coefficient | Standard error | N | R-squared |
|---|---|---|---|---|
| | (1) | (2) | (3) | (4) |
| **Panel A. Selective retention** | | | | |
| **Panel A1** | | | | |
| Sample | Enrolled in the baseline | | | |
| Independent variable | Enrolled in the first follow-up | | | |
| Standardized health status (baseline) | 0.030* | (0.015) | 3,112 | 0.004 |
| Standardized health care utilization (baseline) | 0.034 | (0.022) | 2,377 | 0.003 |
| **Panel A2** | | | | |
| Sample | Enrolled in the baseline | | | |
| Independent variable | Enrolled in the second follow-up | | | |
| Standardized health status (baseline) | 0.044* | (0.024) | 2,681 | 0.010 |
| Standardized health care utilization (baseline) | 0.046 | (0.032) | 2,044 | 0.007 |
| **Panel A3** | | | | |
| Sample | Enrolled in the first follow-up | | | |
| Independent variable | Enrolled in the second follow-up | | | |
| Standardized health status (first follow-up) | 0.021*** | (0.006) | 8,212 | 0.006 |
| Standardized health care utilization (first follow-up) | 0.030*** | (0.008) | 6,255 | 0.004 |
| **Panel B. Selective new enrollment** | | | | |
| **Panel B1** | | | | |
| Sample | Not Enrolled in the baseline | | | |
| Independent variable | Enrolled in the first follow-up | | | |
| Standardized health status (baseline) | -0.002 | (0.007) | 13,226 | 0.000 |
| Standardized health care utilization (baseline) | -0.003 | (0.007) | 10,114 | 0.000 |
| **Panel B2** | | | | |
| Sample | Not Enrolled in the baseline | | | |
| Independent variable | Enrolled in the second follow-up | | | |
| Standardized health status (baseline) | 0.013 | (0.008) | 10,849 | 0.002 |
| Standardized health care utilization (baseline) | 0.017 | (0.012) | 8,276 | 0.002 |
| **Panel B3** | | | | |
| Sample | Not Enrolled in the first follow-up | | | |
| Independent variable | Enrolled in the second follow-up | | | |
| Standardized health status (first follow-up) | 0.004 | (0.009) | 6,248 | 0.000 |
| Standardized health care utilization (first follow-up) | 0.049** | (0.019) | 4,722 | 0.012 |

Notes: This table reports estimation results of running univariate regression of each standardized health characteristics on an enrollment indicator in short and long-run. Panel A examines selective retention. Samples in Panels A1 and A2 are restricted to those who enrolled in the baseline, while Panel A3 restricted to those enrolled in the short run. Panel B examines selective new enrollment. Samples in Panels B1 and B2 are restricted to those who did *not* enroll in the baseline, while Panel B3 restricted to those who did *not* enroll in the short run. Robust standard errors clustered at community level reported in parentheses. *, **, and *** denote statistical significance at 10 %, 5 %, and 1 % level respectively.

Table A6: Additional Results of Selection by Subsidy (No Subsidy vs Subsidy): Characteristics of Compliers, Always Takers, and Never Takers

| | Short run | | | | | | Long run | | | | | |
|---|---|---|---|---|---|---|---|---|---|---|---|---|
| | Mean | | | t-stat | | | Mean | | | t-stat | | |
| | Complier | Always | Never | C=A | C=N | A=N | Complier | Always | Never | C=A | C=N | A=N |
| | (1) | (2) | (3) | (4) | (5) | (6) | (7) | (8) | (9) | (10) | (11) | (12) |
| **Panel A: Health status and health care utilization** | | | | | | | | | | | | |
| **Health status** | | | | | | | | | | | | |
| Illness | | | | | | | | | | | | |
| Number of sick days in the last four weeks | 1.03 | 0.75 | 0.89 | 1.62 | 0.91 | -0.58 | 0.85 | 1.05 | 0.93 | -0.73 | -0.57 | 0.38 |
| Could not do normal activities in the last four weeks | 0.10 | 0.04 | 0.08 | 5.80 | 1.93 | -2.28 | 0.10 | 0.10 | 0.07 | 0.23 | 3.01 | 1.04 |
| No. of days could not perform normal activities in the last four weeks | 0.74 | 0.33 | 0.43 | 3.39 | 3.96 | -0.69 | 0.42 | 0.68 | 0.57 | -1.19 | -1.40 | 0.47 |
| Illness due to Malaria | | | | | | | | | | | | |
| Number of sick days in the last four weeks | 0.29 | 0.32 | 0.16 | -0.30 | 3.12 | 1.63 | 0.34 | 0.53 | 0.14 | -1.05 | 4.15 | 2.05 |
| Could not do normal activities in the last four weeks | 0.04 | 0.02 | 0.02 | 3.23 | 2.16 | -0.71 | 0.07 | 0.03 | 0.01 | 3.65 | 14.84 | 1.29 |
| No. of days could not perform normal activities in the last four weeks | 0.17 | 0.16 | 0.13 | 0.16 | 1.05 | 0.40 | 0.30 | 0.31 | 0.05 | -0.10 | 13.67 | 1.51 |
| **Health care utilization** | | | | | | | | | | | | |
| Visited health facility in the last four weeks | 0.04 | 0.05 | 0.03 | -0.43 | 1.52 | 1.14 | 0.04 | 0.06 | 0.03 | -1.31 | 1.52 | 1.74 |
| Visited health facility in the last six months | 0.07 | 0.09 | 0.06 | -1.16 | 0.73 | 1.37 | 0.05 | 0.13 | 0.05 | -3.40 | -0.28 | 3.12 |
| Number of visits in the last four weeks | 0.06 | 0.08 | 0.05 | -0.80 | 0.81 | 1.07 | 0.02 | 0.10 | 0.06 | -2.25 | -2.79 | 0.99 |
| Visited health facility in the last four weeks for malaria treatment | 0.01 | 0.01 | 0.01 | 0.43 | 1.73 | 0.50 | 0.01 | 0.03 | 0.00 | -1.19 | 8.29 | 2.13 |
| Made out of pocket expense in the last six months | 0.15 | 0.12 | 0.14 | 1.55 | 0.89 | -0.57 | 0.15 | 0.16 | 0.13 | -0.20 | 1.99 | 1.03 |
| **Panel B: Other characteristics** | | | | | | | | | | | | |
| Age | 24.34 | 20.48 | 24.39 | 3.61 | -0.05 | -2.58 | 18.90 | 21.46 | 27.08 | -1.80 | -10.13 | -3.43 |
| Male | 0.51 | 0.47 | 0.48 | 1.18 | 1.17 | -0.15 | 0.51 | 0.44 | 0.51 | 1.91 | -0.03 | -1.71 |
| Christian | 0.43 | 0.51 | 0.40 | -2.74 | 1.24 | 2.90 | 0.37 | 0.53 | 0.42 | -4.59 | -2.40 | 3.00 |
| Dagaaba (ethnic group) | 0.61 | 0.54 | 0.44 | 2.14 | 6.89 | 2.79 | 0.59 | 0.53 | 0.51 | 1.70 | 4.39 | 0.51 |
| Has some formal education | 0.38 | 0.35 | 0.26 | 0.81 | 5.61 | 2.78 | 0.42 | 0.35 | 0.30 | 2.16 | 7.62 | 1.43 |
| Household size | 9.35 | 8.34 | 8.57 | 5.11 | 5.41 | -0.92 | 10.19 | 7.99 | 8.86 | 13.02 | 12.33 | -4.07 |
| Household assets (principal component score) | 1.09 | 0.77 | 0.59 | 2.56 | 5.69 | 1.19 | 2.29 | 0.16 | 0.59 | 23.27 | 71.53 | -3.50 |
| Ever enrolled in NHIS | 0.30 | 0.65 | 0.30 | -12.98 | 0.23 | 10.02 | 0.40 | 0.41 | 0.40 | -0.22 | -0.25 | 0.08 |
| Currently enrolled in NHIS | 0.05 | 0.47 | 0.15 | -14.99 | -6.38 | 9.39 | 0.14 | 0.30 | 0.17 | -5.10 | -2.36 | 3.79 |

Note: This table presents the mean characteristics of compliers, always takers, and never takers comparing control group with any subsidy group, which are estimated from Equation (3). Panel A presents statistics on the component variables of standardized health status and health care utilization. Panel B presents statistics on the other socio-economic characteristics. Columns 4-6 and 10-12 present the t-statistics from the two-sample t-test comparing compliers with always takers, compliers with never takers, and always takers with never takers, respectively.

Table A7: Additional Results of Selection by Subsidy (No Subsidy vs Partial Subsidy): Characteristics of Compliers, Always Takers, and Never Takers

| | Short run | | | | | | Long run | | | | | |
|---|---|---|---|---|---|---|---|---|---|---|---|---|
| | Mean | | | t-stat | | | Mean | | | t-stat | | |
| | Complier | Always | Never | C=A | C=N | A=N | Complier | Always | Never | C=A | C=N | A=N |
| | (1) | (2) | (3) | (4) | (5) | (6) | (7) | (8) | (9) | (10) | (11) | (12) |
| **Panel A: Health status and health care utilization** | | | | | | | | | | | | |
| **Health status** | | | | | | | | | | | | |
| Illness | | | | | | | | | | | | |
| Number of sick days in the last four weeks | 1.34 | 0.75 | 1.15 | 3.43 | 0.68 | -1.28 | 1.18 | 1.05 | 1.08 | 0.47 | 0.43 | -0.07 |
| Could not do normal activities in the last four weeks | 0.10 | 0.04 | 0.09 | 5.67 | 0.58 | -2.30 | 0.11 | 0.10 | 0.07 | 0.89 | 3.36 | 1.08 |
| No. of days could not perform normal activities in the last four weeks | 0.71 | 0.33 | 0.55 | 3.13 | 1.33 | -1.22 | 0.32 | 0.68 | 0.49 | -1.64 | -1.31 | 0.74 |
| Illness due to Malaria | | | | | | | | | | | | |
| Number of sick days in the last four weeks | 0.43 | 0.32 | 0.17 | 1.29 | 4.41 | 1.36 | 0.51 | 0.53 | 0.20 | -0.14 | 3.20 | 1.59 |
| Could not do normal activities in the last four weeks | 0.04 | 0.02 | 0.03 | 3.47 | 1.30 | -0.82 | 0.07 | 0.03 | 0.02 | 3.52 | 8.17 | 0.93 |
| No. of days could not perform normal activities in the last four weeks | 0.14 | 0.16 | 0.14 | -0.22 | 0.12 | 0.25 | 0.23 | 0.32 | 0.06 | -0.51 | 5.89 | 1.44 |
| **Health care utilization** | | | | | | | | | | | | |
| Visited health facility in the last four weeks | 0.08 | 0.05 | 0.04 | 2.13 | 2.38 | 0.15 | 0.07 | 0.06 | 0.05 | 0.51 | 1.95 | 0.64 |
| Visited health facility in the last six months | 0.09 | 0.09 | 0.07 | 0.15 | 1.00 | 0.62 | 0.08 | 0.13 | 0.06 | -2.41 | 1.15 | 2.65 |
| Number of visits in the last four weeks | 0.13 | 0.08 | 0.07 | 1.62 | 2.39 | 0.27 | 0.05 | 0.10 | 0.11 | -1.28 | -1.70 | -0.08 |
| Visited health facility in the last four weeks for malaria treatment | 0.01 | 0.01 | 0.01 | 0.73 | 0.65 | -0.04 | 0.02 | 0.03 | 0.00 | -0.55 | 6.09 | 1.96 |
| Made out of pocket expense in the last six months | 0.19 | 0.12 | 0.16 | 3.75 | 1.14 | -1.36 | 0.27 | 0.16 | 0.14 | 4.16 | 7.11 | 0.67 |
| **Panel B: Other characteristics** | | | | | | | | | | | | |
| Age | 26.20 | 20.48 | 25.19 | 5.38 | 0.66 | -2.50 | 26.07 | 21.29 | 26.39 | 3.39 | -0.26 | -2.73 |
| Male | 0.50 | 0.47 | 0.44 | 0.83 | 1.68 | 0.73 | 0.48 | 0.44 | 0.50 | 1.19 | -0.64 | -1.33 |
| Christian | 0.47 | 0.51 | 0.44 | -1.41 | 0.97 | 1.66 | 0.49 | 0.54 | 0.43 | -1.27 | 2.24 | 2.35 |
| Dagaaba (ethnic group) | 0.79 | 0.54 | 0.50 | 8.72 | 8.73 | 0.96 | 0.80 | 0.53 | 0.62 | 8.01 | 7.80 | -1.93 |
| Has some formal education | 0.34 | 0.35 | 0.26 | -0.33 | 2.79 | 2.26 | 0.36 | 0.35 | 0.29 | 0.16 | 2.72 | 1.45 |
| Household size | 8.50 | 8.34 | 8.16 | 0.78 | 1.86 | 0.69 | 8.23 | 7.99 | 8.68 | 1.32 | -2.80 | -2.85 |
| Household assets (principal component score) | 0.81 | 0.77 | 0.41 | 0.32 | 3.64 | 2.16 | 2.08 | 0.17 | 0.51 | 22.86 | 23.47 | -2.39 |
| Ever enrolled in NHIS | 0.36 | 0.65 | 0.29 | -10.82 | 2.25 | 8.71 | 0.50 | 0.41 | 0.41 | 2.67 | 3.43 | -0.08 |
| Currently enrolled in NHIS | 0.16 | 0.47 | 0.10 | -10.98 | 3.06 | 10.43 | 0.26 | 0.30 | 0.19 | -1.44 | 2.97 | 2.82 |

Note: This table presents the mean characteristics of compliers, always takers, and never takers comparing control group with partial subsidy group, which are estimated from Equation (3). Panel A presents statistics on the component variables of standardized health status and health care utilization. Panel B presents statistics on the other socio-economic characteristics. Columns 4-6 and 10-12 present the t-statistics from the two-sample t-test comparing compliers with always takers, compliers with never takers, and always takers with never takers, respectively.

Table A8: Additional Results of Selection by Subsidy (No Subsidy vs Full Subsidy): Characteristics of Compliers, Always Takers, and Never Takers

| | Short run | | | | | | Long run | | | | | |
|---|---|---|---|---|---|---|---|---|---|---|---|---|
| | Mean | | | t-stat | | | Mean | | | t-stat | | |
| | Complier | Always | Never | C=A | C=N | A=N | Complier | Always | Never | C=A | C=N | A=N |
| | (1) | (2) | (3) | (4) | (5) | (6) | (7) | (8) | (9) | (10) | (11) | (12) |
| **Panel A: Health status and health care utilization** | | | | | | | | | | | | |
| **Health status** | | | | | | | | | | | | |
| Illness | | | | | | | | | | | | |
| Number of sick days in the last four weeks | 0.79 | 0.75 | 0.59 | 0.21 | 1.37 | 0.73 | 0.52 | 1.05 | 0.79 | -2.00 | -1.60 | 0.84 |
| Could not do normal activities in the last four weeks | 0.11 | 0.04 | 0.07 | 5.93 | 2.30 | -1.24 | 0.09 | 0.10 | 0.07 | -0.28 | 1.23 | 0.90 |
| No. of days could not perform normal activities in the last four weeks | 0.77 | 0.33 | 0.31 | 3.59 | 5.09 | 0.17 | 0.48 | 0.68 | 0.58 | -0.94 | -0.66 | 0.41 |
| Illness due to Malaria | | | | | | | | | | | | |
| Number of sick days in the last four weeks | 0.18 | 0.32 | 0.14 | -1.54 | 0.67 | 1.64 | 0.22 | 0.53 | 0.09 | -1.69 | 4.13 | 2.36 |
| Could not do normal activities in the last four weeks | 0.04 | 0.02 | 0.02 | 3.04 | 1.87 | -0.29 | 0.07 | 0.03 | 0.01 | 3.46 | 11.66 | 1.46 |
| No. of days could not perform normal activities in the last four weeks | 0.19 | 0.16 | 0.12 | 0.46 | 1.21 | 0.48 | 0.33 | 0.32 | 0.04 | 0.06 | 11.63 | 1.54 |
| **Health care utilization** | | | | | | | | | | | | |
| Visited health facility in the last four weeks | 0.01 | 0.05 | 0.01 | -2.45 | 0.80 | 2.56 | 0.02 | 0.06 | 0.01 | -2.38 | 3.06 | 2.99 |
| Visited health facility in the last six months | 0.05 | 0.09 | 0.05 | -2.19 | 0.38 | 1.87 | 0.04 | 0.13 | 0.05 | -3.92 | -0.72 | 3.25 |
| Number of visits in the last four weeks | 0.00 | 0.08 | 0.01 | -2.72 | -1.54 | 2.20 | 0.00 | 0.10 | 0.01 | -2.81 | -2.36 | 2.33 |
| Visited health facility in the last four weeks for malaria treatment | 0.01 | 0.01 | 0.00 | 0.20 | N/A | 1.42 | 0.01 | 0.03 | 0.00 | -1.53 | N/A | 2.26 |
| Made out of pocket expense in the last six months | 0.12 | 0.12 | 0.11 | -0.17 | 0.64 | 0.59 | 0.08 | 0.16 | 0.12 | -3.18 | -2.72 | 1.17 |
| **Panel B: Other characteristics** | | | | | | | | | | | | |
| Age | 22.88 | 20.48 | 23.49 | 2.24 | -0.42 | -1.66 | 14.60 | 21.29 | 27.62 | -4.75 | -12.06 | -3.53 |
| Male | 0.51 | 0.47 | 0.52 | 1.47 | -0.17 | -1.05 | 0.51 | 0.44 | 0.52 | 1.85 | -0.43 | -1.74 |
| Christian | 0.40 | 0.51 | 0.36 | -3.78 | 1.17 | 3.33 | 0.28 | 0.54 | 0.41 | -7.39 | -5.20 | 2.95 |
| Dagaaba (ethnic group) | 0.46 | 0.54 | 0.37 | -2.91 | 2.67 | 3.91 | 0.45 | 0.53 | 0.41 | -2.32 | 1.43 | 2.71 |
| Has some formal education | 0.40 | 0.35 | 0.25 | 1.71 | 4.89 | 2.50 | 0.48 | 0.35 | 0.30 | 3.70 | 7.46 | 1.24 |
| Household size | 10.02 | 8.34 | 9.04 | 8.58 | 4.41 | -2.30 | 11.44 | 7.99 | 9.05 | 22.92 | 15.08 | -4.09 |
| Household assets (principal component score) | 1.30 | 0.77 | 0.79 | 4.33 | 3.73 | -0.10 | 2.44 | 0.17 | 0.71 | 26.43 | 38.76 | -4.06 |
| Ever enrolled in NHIS | 0.26 | 0.65 | 0.30 | -14.91 | -1.41 | 8.23 | 0.33 | 0.41 | 0.41 | -2.12 | -2.85 | 0.04 |
| Currently enrolled in NHIS | -0.04 | 0.47 | 0.21 | -18.58 | -9.02 | 6.35 | 0.06 | 0.30 | 0.15 | -7.80 | -5.38 | 4.19 |

Note: This table presents the mean characteristics of compliers, always takers, and never takers comparing control group with full subsidy group, which are estimated from Equation (3). Panel A presents statistics on the component variables of standardized health status and health care utilization. Panel B presents statistics on the other socio-economic characteristics. Columns 4-6 and 10-12 present the t-statistics from the two-sample t-test comparing compliers with always takers, compliers with never takers, and always takers with never takers, respectively.

Table A9: Results of Selection by Subsidy Level (Partial vs Full): Baseline Health Characteristics

| | Baseline health status | | | | | | Baseline health care utilization | | | | |
|---|---|---|---|---|---|---|---|---|---|---|---|
| | # Days ill last month | Could not perform normal daily activities due to illness last month | # days could not perform normal daily activities in the last month | # Days ill last month due to malaria | Could not perform normal daily activities due to illness last month due to malaria | # days could not perform normal daily activities in the last month due to malaria | Visited health facility in last four weeks | Visited health facility in last six months | # of visits in last four weekss | Visited facility for malaria treatment in the last four weeks | Made out-of-pocket for health service in the last six months |
| | (1) | (2) | (3) | (4) | (5) | (6) | (7) | (8) | (9) | (10) | (11) |
| Sample | Enrolled in the short run | | | | | | | | | | |
| **Panel A** | | | | | | | | | | | |
| Full subsidy | -0.196 | 0.026 | 0.288 | 0.004 | 0.018 | 0.135 | -0.037 | -0.014 | -0.093* | 0.010 | -0.033 |
| | (0.494) | (0.030) | (0.342) | (0.242) | (0.020) | (0.187) | (0.032) | (0.031) | (0.051) | (0.014) | (0.036) |
| N | 1,238 | 1,233 | 1,186 | 1,238 | 1,233 | 1,186 | 1,090 | 1,244 | 1,091 | 1,090 | 1,244 |
| R-squared | 0.004 | 0.006 | 0.005 | 0.009 | 0.013 | 0.005 | 0.023 | 0.008 | 0.018 | 0.012 | 0.007 |
| Sample | Enrolled in the long run | | | | | | | | | | |
| **Panel B** | | | | | | | | | | | |
| Full subsidy | -0.097 | 0.022 | 0.315 | 0.154 | 0.040 | 0.262 | 0.013 | 0.020 | 0.025 | 0.021 | -0.045 |
| | (0.437) | (0.042) | (0.393) | (0.410) | (0.042) | (0.384) | (0.042) | (0.038) | (0.051) | (0.026) | (0.062) |
| N | 661 | 657 | 625 | 661 | 657 | 625 | 572 | 664 | 573 | 572 | 664 |
| R-squared | 0.010 | 0.015 | 0.009 | 0.018 | 0.025 | 0.011 | 0.033 | 0.020 | 0.033 | 0.027 | 0.034 |
| Sample | Enrolled in the short and long run | | | | | | | | | | |
| **Panel C** | | | | | | | | | | | |
| Full subsidy | 0.940 | 0.089 | 1.024 | 0.775 | 0.080 | 0.776 | -0.053 | 0.015 | -0.046 | 0.003 | 0.083* |
| | (0.652) | (0.061) | (0.642) | (0.651) | (0.063) | (0.640) | (0.048) | (0.048) | (0.048) | (0.021) | (0.047) |
| N | 422 | 421 | 408 | 422 | 421 | 408 | 375 | 424 | 375 | 375 | 424 |
| R-squared | 0.038 | 0.033 | 0.049 | 0.056 | 0.053 | 0.057 | 0.057 | 0.016 | 0.052 | 0.008 | 0.038 |
| Sample | Enrolled in all rounds | | | | | | | | | | |
| **Panel D** | | | | | | | | | | | |
| Full subsidy | -1.043 | -0.018 | 0.202 | -0.336 | 0.057 | 0.335 | 0.105 | 0.070 | 0.063 | 0.072 | -0.125 |
| | (0.846) | (0.105) | (0.446) | (0.528) | (0.058) | (0.282) | (0.079) | (0.086) | (0.102) | (0.056) | (0.104) |
| N | 127 | 126 | 117 | 127 | 126 | 117 | 108 | 127 | 109 | 108 | 127 |
| R-squared | 0.032 | 0.037 | 0.006 | 0.025 | 0.049 | 0.018 | 0.114 | 0.065 | 0.141 | 0.076 | 0.051 |

Note: This table presents selection by subsidy level based on the component variables of standardized health status and health care utilization.

Table A10: Results of Selection by Subsidy Level (Partial vs Full): Short and Long-run Health Care Utilization

| | Short-run health care utilization | | | | | Long-run health care utilization | | | | |
|---|---|---|---|---|---|---|---|---|---|---|
| | Visited health facility in last four weeks | Visited health facility in last six months | # of visits in last four weekss | Visited facility for malaria treatment in the last four weeks | Made out-of-pocket for health service in the last six months | Visited health facility in last four weeks | Visited health facility in last six months | # of visits in last four weekss | Visited facility for malaria treatment in the last four weeks | Made out-of-pocket for health service in the last six months |
| | (1) | (2) | (3) | (4) | (5) | (6) | (7) | (8) | (9) | (10) |
| Sample | Enrolled in the short run | | | | | | | | | |
| **Panel A** | | | | | | | | | | |
| Full subsidy | -0.010 | -0.001 | 0.004 | -0.018 | -0.006 | | | | | |
| | (0.017) | (0.027) | (0.012) | (0.011) | (0.016) | | | | | |
| N | 1,152 | 1,223 | 1,148 | 1,200 | 1,244 | | | | | |
| R-squared | 0.017 | 0.025 | 0.010 | 0.008 | 0.008 | | | | | |
| Sample | Enrolled in the long run | | | | | | | | | |
| **Panel B** | | | | | | | | | | |
| Full subsidy | 0.005 | -0.022 | 0.030 | -0.032** | -0.019 | -0.122*** | -0.245*** | -0.101** | -0.091** | -0.032 |
| | (0.033) | (0.032) | (0.043) | (0.012) | (0.018) | (0.035) | (0.076) | (0.041) | (0.037) | (0.023) |
| N | 611 | 651 | 608 | 635 | 664 | 664 | 664 | 664 | 664 | 664 |
| R-squared | 0.009 | 0.022 | 0.009 | 0.008 | 0.009 | 0.045 | 0.091 | 0.034 | 0.035 | 0.018 |
| Sample | Enrolled in the short and long run | | | | | | | | | |
| **Panel C** | | | | | | | | | | |
| Full subsidy | -0.015 | 0.053 | 0.014 | -0.010 | -0.022 | -0.102** | -0.197*** | -0.074* | -0.081** | -0.013 |
| | (0.029) | (0.054) | (0.044) | (0.025) | (0.015) | (0.037) | (0.052) | (0.039) | (0.037) | (0.023) |
| N | 393 | 420 | 392 | 409 | 424 | 424 | 424 | 424 | 424 | 424 |
| R-squared | 0.007 | 0.011 | 0.009 | 0.002 | 0.005 | 0.035 | 0.097 | 0.024 | 0.033 | 0.005 |
| Sample | Enrolled in all rounds | | | | | | | | | |
| **Panel D** | | | | | | | | | | |
| Full subsidy | -0.001 | -0.064 | -0.011 | -0.068* | -0.010 | -0.170*** | -0.133 | -0.087 | -0.093 | -0.075* |
| | (0.035) | (0.096) | (0.033) | (0.037) | (0.018) | (0.055) | (0.108) | (0.076) | (0.055) | (0.036) |
| N | 120 | 124 | 119 | 124 | 127 | 127 | 127 | 127 | 127 | 127 |
| R-squared | 0.059 | 0.154 | 0.096 | 0.051 | 0.014 | 0.071 | 0.074 | 0.029 | 0.042 | 0.056 |

Note: This table presents selection by subsidy level based on the component variables of standardized health care utilization in short and long run.

Table A11: Effects on Health Status (Short Run)

| | Short run | | | | |
|---|---|---|---|---|---|
| | Feeling Unhealthy | # of days ill last four weeks | Could not perform normal daily activities due to illness last four weeks | # of days could not perform normal daily activities due to illness in the last four weeks | Standardized treatment effects |
| | (1) | (2) | (3) | (4) | (5) |
| **Panel A: ITT results** | | | | | |
| **Panel A1** | | | | | |
| Any subsidy | -0.126*** | -0.337** | -0.014 | -0.170 | -0.029** |
| | (0.037) | (0.142) | (0.019) | (0.392) | (0.014) |
| | [0.003] | [0.035] | [0.535] | [0.696] | |
| | {0.035} | {0.134} | {0.677} | {0.723} | |
| R-squared | 0.192 | 0.086 | 0.080 | 0.093 | 0.063 |
| **Panel A2** | | | | | |
| Partial subsidy | -0.129*** | -0.316** | -0.013 | -0.081 | -0.026* |
| | (0.037) | (0.135) | (0.018) | (0.370) | (0.013) |
| | [0.003] | [0.039] | [0.558] | [0.858] | |
| | {0.035} | {0.136} | {0.704} | {0.862} | |
| Full subsidy | -0.117** | -0.417* | -0.021 | -0.516 | -0.042* |
| | (0.044) | (0.210) | (0.030) | (0.512) | (0.022) |
| | [0.010] | [0.079] | [0.525] | [0.360] | |
| | {0.106} | {0.212} | {0.540} | {0.489} | |
| R-squared | 0.192 | 0.086 | 0.080 | 0.094 | 0.064 |
| **Panel A3** | | | | | |
| 1/3 subsidy | -0.118*** | -0.412** | -0.012 | -0.392 | -0.035* |
| | (0.042) | (0.166) | (0.024) | (0.418) | (0.018) |
| | [0.011] | [0.036] | [0.667] | [0.406] | |
| | {0.081} | {0.100} | {0.669} | {0.544} | |
| 2/3 subsidy | -0.137*** | -0.244 | -0.013 | 0.165 | -0.018 |
| | (0.044) | (0.171) | (0.020) | (0.428) | (0.015) |
| | [0.011] | [0.235] | [0.620] | [0.775] | |
| | {0.088} | {0.520} | {0.735} | {0.759} | |
| Full subsidy | -0.119*** | -0.402* | -0.021 | -0.458 | -0.040* |
| | (0.044) | (0.213) | (0.030) | (0.511) | (0.022) |
| | [0.009] | [0.111] | [0.534] | [0.424] | |
| | {0.105} | {0.248} | {0.554} | {0.554} | |
| R-squared | 0.193 | 0.087 | 0.080 | 0.095 | 0.064 |
| Number of observations | 861 | 2,768 | 2,775 | 2,677 | 8,824 |
| Control group mean | 0.817 | 0.617 | 0.081 | 1.379 | -0.019 |
| **P-values on test of equality:** | | | | | |
| Partial subsidy = Full subsidy | 0.737 | 0.473 | 0.712 | 0.180 | 0.287 |
| 1/3 subsidy = 2/3 subsidy | 0.685 | 0.395 | 0.942 | 0.208 | 0.392 |
| 1/3 subsidy = Full subsidy | 0.984 | 0.954 | 0.713 | 0.858 | 0.755 |
| 2/3 subsidy = Full subsidy | 0.662 | 0.331 | 0.761 | 0.102 | 0.229 |

Notes: Panel A reports ITT results. Panels A1, A2, and A3 report the effects of receiving any subsidy, partial and full subsidy, and each subsidy level (1/3, 2/3, and full) on health status in the short run. All regressions include a standard set of covariates (individual, household, and community) and baseline measure of dependent variable. Standardized treatment effects are reported in Column 5. *P*-values for the equality of effect estimates for various pairs of treatment groups are also presented. Robust standard errors clustered at community level are reported in parentheses. Wild-cluster bootstrap-t *p*-values are reported in square brackets. Family-wise *p*-values are reported in curly brackets. *, **, and *** denote statistical significance at 10 %, 5 %, and 1 % levels, respectively.

Table A12: Effects on Health Status (Long Run)

| | Long run | | | | |
|---|---|---|---|---|---|
| | Feeling Unhealthy | # of days ill last four weeks | Could not perform normal daily activities due to illness last four weeks | # of days could not perform normal daily activities due to illness in the last four weeks | Standardized treatment effects |
| | (1) | (2) | (3) | (4) | (5) |
| **Panel A: ITT results** | | | | | |
| **Panel A1** | | | | | |
| Any subsidy | 0.158*** | 0.148 | 0.035** | 0.181** | 0.039** |
| | (0.054) | (0.094) | (0.013) | (0.088) | (0.015) |
| | [0.042] | [0.183] | [0.039] | [0.121] | |
| | {0.088} | {0.179} | {0.088} | {0.135} | |
| R-squared | 0.308 | 0.072 | 0.091 | 0.067 | 0.057 |
| **Panel A2** | | | | | |
| Partial subsidy | 0.163*** | 0.286*** | 0.047*** | 0.279*** | 0.057*** |
| | (0.056) | (0.084) | (0.013) | (0.079) | (0.013) |
| | [0.029] | [0.022] | [0.006] | [0.020] | |
| | {0.036} | {0.031} | {0.031} | {0.031} | |
| Full subsidy | 0.138 | -0.362** | -0.008 | -0.185 | -0.029 |
| | (0.089) | (0.171) | (0.020) | (0.182) | (0.027) |
| | [0.222] | [0.105] | [0.785] | [0.521] | |
| | {0.503} | {0.349} | {0.781} | {0.586} | |
| R-squared | 0.308 | 0.084 | 0.100 | 0.078 | 0.068 |
| **Panel A3** | | | | | |
| 1/3 subsidy | 0.092 | 0.191 | 0.033** | 0.231* | 0.041** |
| | (0.066) | (0.137) | (0.016) | (0.126) | (0.019) |
| | [0.238] | [0.235] | [0.070] | [0.119] | |
| | {0.421} | {0.421} | {0.312} | {0.331} | |
| 2/3 subsidy | 0.224*** | 0.358*** | 0.057*** | 0.316*** | 0.070*** |
| | (0.072) | (0.099) | (0.016) | (0.105) | (0.018) |
| | [0.017] | [0.009] | [0.016] | [0.032] | |
| | {0.042} | {0.033} | {0.034} | {0.042} | |
| Full subsidy | 0.148* | -0.348** | -0.005 | -0.177 | -0.026 |
| | (0.087) | (0.165) | (0.020) | (0.173) | (0.026) |
| | [0.170] | [0.093] | [0.817] | [0.456] | |
| | {0.444} | {0.354} | {0.837} | {0.597} | |
| R-squared | 0.314 | 0.085 | 0.102 | 0.078 | 0.069 |
| Number of observations | 658 | 2,666 | 2,661 | 2,564 | 8,309 |
| Control group mean | 0.791 | 0.413 | 0.013 | 0.096 | 0.011 |
| **P-values on test of equality:** | | | | | |
| Partial subsidy = Full subsidy | 0.773 | 0.000 | 0.008 | 0.013 | 0.003 |
| 1/3 subsidy = 2/3 subsidy | 0.161 | 0.319 | 0.236 | 0.616 | 0.257 |
| 1/3 subsidy = Full subsidy | 0.418 | 0.008 | 0.090 | 0.014 | 0.015 |
| 2/3 subsidy = Full subsidy | 0.499 | 0.000 | 0.006 | 0.025 | 0.003 |

Notes: Panel A reports ITT results. Panels A1, A2, and A3 report the effects of receiving any subsidy, partial and full subsidy, and each subsidy level (1/3, 2/3, and full) on health status in the long run. All regressions include a standard set of covariates (individual, household, and community) and baseline measure of dependent variable. Standardized treatment effects are reported in Column 5. *P*-values for the equality of effect estimates for various pairs of treatment groups are also presented. Robust standard errors clustered at community level are reported in parentheses. Wild-cluster bootstrap-t *p*-values are reported in square brackets. Family-wise *p*-values are reported in curly brackets. *, **, and *** denote statistical significance at 10 %, 5 %, and 1 % levels, respectively.

Table A13: Effects on Health Behaviors

| | Short run | Long run | | | |
|---|---|---|---|---|---|
| | Sleep under mosquito nets | Have mosquito nets | Sleep under mosquito nets | Water safe to drink | Standardized treatment effects |
| | (1) | (2) | (3) | (4) | (5) |
| **Panel A: ITT results** | | | | | |
| **Panel A1** | | | | | |
| Any subsidy | 0.123 | -0.028 | 0.023 | -0.057 | -0.037 |
| | (0.104) | (0.096) | (0.121) | (0.048) | (0.046) |
| | [0.251] | [0.839] | [0.875] | [0.289] | |
| | | {0.933} | {0.933} | {0.588} | |
| R-squared | 0.233 | 0.258 | 0.235 | 0.257 | 0.172 |
| **Panel A2** | | | | | |
| Partial subsidy | 0.098 | 0.039 | 0.036 | -0.071 | -0.018 |
| | (0.113) | (0.094) | (0.123) | (0.045) | (0.044) |
| | [0.457] | [0.758] | [0.806] | [0.149] | |
| | | {0.926} | {0.926} | {0.522} | |
| Full subsidy | 0.227* | -0.269** | -0.044 | -0.014 | -0.117** |
| | (0.118) | (0.106) | (0.118) | (0.068) | (0.051) |
| | [0.113] | [0.064] | [0.749] | [0.878] | |
| | | {0.241} | {0.946} | {0.946} | |
| R-squared | 0.247 | 0.318 | 0.259 | 0.275 | 0.179 |
| **Panel A3** | | | | | |
| 1/3 subsidy | 0.020 | 0.146 | 0.072 | -0.054 | 0.021 |
| | (0.110) | (0.117) | (0.127) | (0.057) | (0.047) |
| | [0.886] | [0.343] | [0.664] | [0.394] | |
| | | {0.716} | {0.724} | {0.724} | |
| 2/3 subsidy | 0.158 | -0.065 | 0.009 | -0.087* | -0.055 |
| | (0.141) | (0.089) | (0.131) | (0.044) | (0.049) |
| | [0.396] | [0.567] | [0.959] | [0.061] | |
| | | {0.796} | {0.956} | {0.355} | |
| Full subsidy | 0.238** | -0.294*** | -0.050 | -0.017 | -0.127** |
| | (0.118) | (0.097) | (0.120) | (0.068) | (0.051) |
| | [0.088] | [0.022] | [0.723] | [0.829] | |
| | | {0.143} | {0.931} | {0.931} | |
| R-squared | 0.252 | 0.333 | 0.260 | 0.276 | 0.182 |
| Number of observations | 1,422 | 1,101 | 1,092 | 497 | 2,069 |
| Control group mean | 0.447 | 0.290 | 0.661 | 0.080 | 0.007 |
| **P-values on test of equality:** | | | | | |
| Partial subsidy = Full subsidy | 0.274 | 0.001 | 0.179 | 0.166 | 0.003 |
| 1/3 subsidy = 2/3 subsidy | 0.303 | 0.043 | 0.382 | 0.482 | 0.008 |
| 1/3 subsidy = Full subsidy | 0.096 | 0.0001 | 0.131 | 0.490 | 0.00004 |
| 2/3 subsidy = Full subsidy | 0.544 | 0.011 | 0.340 | 0.106 | 0.049 |

Note: Health behaviors are measured for those aged 12 years and above. Dependent variable in Column 4 is an indicator variable of whether a household member does anything to their water to make it safe to drink. Panel A reports ITT results. Panels A1, A2, and A3 report the effects of receiving any subsidy, partial and full subsidy, and each subsidy level (1/3, 2/3, and full), respectively. All regressions include a standard set of covariates (individual, household, and community) and baseline measure of dependent variable. Standardized treatment effect in the long run is reported in Column 5. *P*-values for the equality of effect estimates for various pairs of treatment groups are also presented. Robust standard errors clustered at community level are reported in parentheses. Wild-cluster bootstrap-t *p*-values are reported in square brackets. Family-wise *p*-values are reported in curly brackets *, **, and *** denote statistical significance at 10 %, 5 %, and 1 % levels, respectively.

**Appendix B**

Figure B.1: Original Study Design

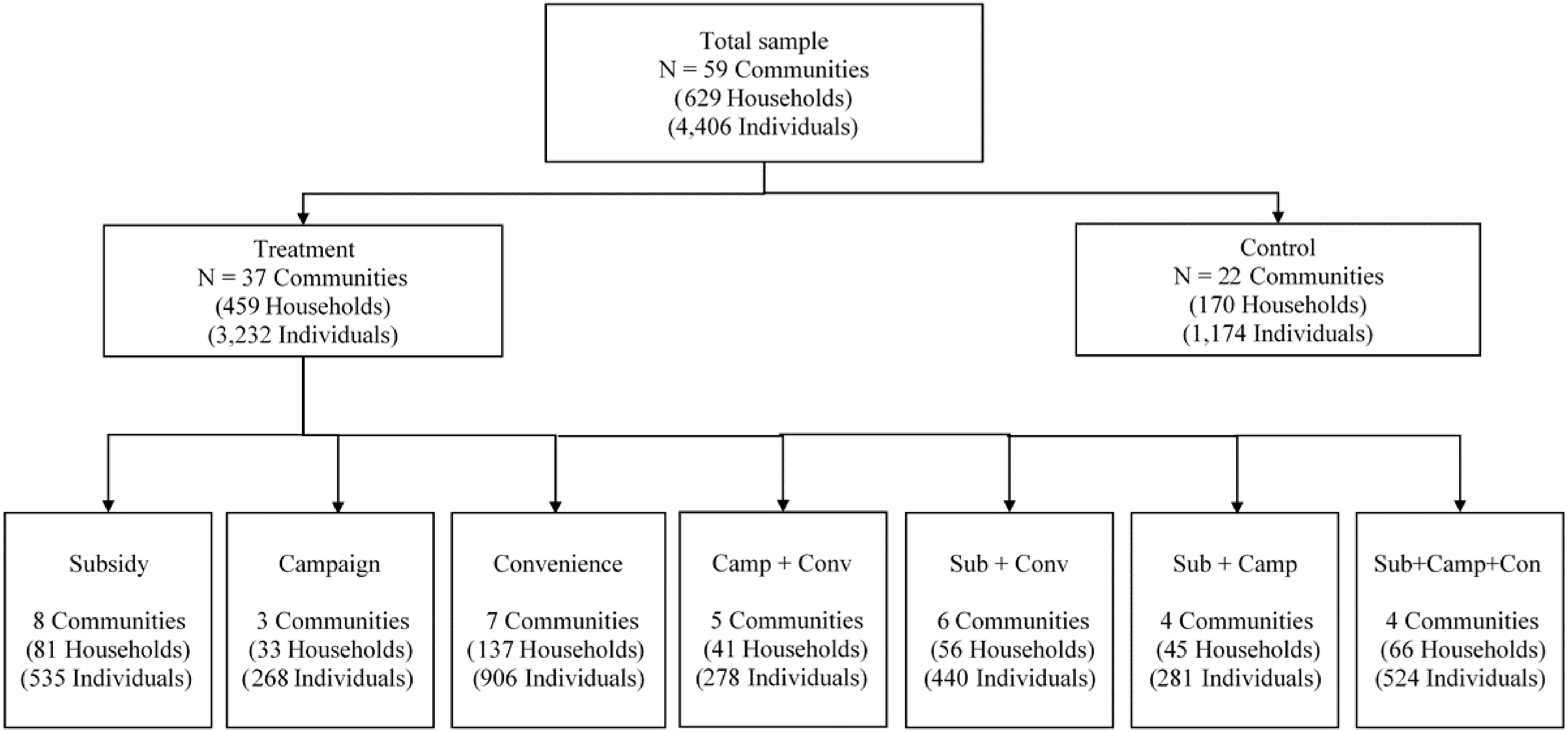

Table B.1: Effects of the Original Interventions on Enrollment

| | Enrollment | |
|---|---|---|
| | Short-run | Long-run |
| | (1) | (2) |
| Subsidy only | 0.436*** | 0.160* |
| | (0.046) | (0.082) |
| Campaign only | 0.161** | 0.044 |
| | (0.080) | (0.066) |
| Convenience only | 0.007 | 0.195*** |
| | (0.066) | (0.072) |
| Campaign & Convenience | 0.231 | 0.182 |
| | (0.165) | (0.159) |
| Subsidy & Convenience | 0.347*** | 0.155** |
| | (0.078) | (0.064) |
| Subsidy & Campaign | 0.520*** | 0.080 |
| | (0.075) | (0.094) |
| Subsidy & Camp & Conven | 0.458*** | 0.397*** |
| | (0.064) | (0.083) |
| R-squared | 0.318 | 0.166 |
| Mean | 0.504 | 0.379 |
| Control group mean | 0.272 | 0.230 |
| Number of observations | 4,168 | 3,415 |
| **P-value on test of equality** | | |
| Sub + Camp = Sub & Camp | 0.477 | 0.330 |
| Sub + Conv = Sub & Conv | 0.323 | 0.090 |
| Camp + Conv = Camp & Conv | 0.756 | 0.770 |
| Sub + Camp + Conv = Sub & Camp & Conv | 0.211 | 0.991 |

Notes: This table presents the effects of original intervention on enrollment in health insurance in short and long run. All regressions include a standard set of covariates (individual, household, and community) and baseline measure of dependent variable. *P*-values for the equality of effect estimates are also presented. Robust standard errors clustered at the community level are reported in parentheses. *, **, and *** denote statistical significance at 10 %, 5 %, and 1 % levels, respectively.

Table B.2: Effects on Enrollment with Restricted Sample

| | Enrollment | |
|---|---|---|
| | Short-run | Long-run |
| | (1) | (4) |
| **Panel A** | | |
| Any Subsidy | 0.424*** | 0.135* |
| | (0.044) | (0.074) |
| R-squared | 0.399 | 0.228 |
| **Panel B** | | |
| Partial subsidy (positive price) | 0.405*** | 0.095 |
| | (0.047) | (0.067) |
| Full subsidy (free) | 0.514*** | 0.317*** |
| | (0.090) | (0.105) |
| R-squared | 0.401 | 0.238 |
| **Panel C** | | |
| 1/3 subsidy | 0.387*** | 0.137* |
| | (0.084) | (0.080) |
| 2/3 subsidy | 0.419*** | 0.063 |
| | (0.062) | (0.067) |
| Full subsidy (free) | 0.514*** | 0.316*** |
| | (0.090) | (0.105) |
| R-squared | 0.401 | 0.239 |
| Mean | 0.405 | 0.290 |
| Control group mean | 0.272 | 0.230 |
| Number of observations | 1,614 | 1,304 |

Notes: This table corresponds to Table 2, but the sample is restricted to subsidy only and control groups. All regressions include a standard set of covariates (individual, household, and community) and baseline measure of dependent variable. Robust standard errors clustered at community level are reported in parentheses. *, **, and *** denote statistical significance at 10 %, 5 %, and 1 % levels, respectively.

Table B.3: Effects on Health Care Utilization with Restricted Sample (Short Run)

| | Short run | | | | | |
|---|---|---|---|---|---|---|
| | Visited health facility in last four weeks | Visited health facility in last six months | # of visits in last four weekss | Visited Facility for malaria treatment in the last four weeks | Made an out-of-pocket for health service in the last six months | Standardized treatment effects |
| | (1) | (2) | (3) | (4) | (5) | (6) |
| **Panel A** | | | | | | |
| Any subsidy | -0.011 | -0.009 | -0.007 | 0.011 | -0.018 | -0.003 |
| | (0.010) | (0.020) | (0.024) | (0.008) | (0.016) | (0.009) |
| R-squared | 0.121 | 0.137 | 0.142 | 0.113 | 0.139 | 0.086 |
| **Panel B** | | | | | | |
| Partial subsidy | -0.012 | -0.005 | -0.004 | 0.014 | -0.011 | 0.001 |
| | (0.012) | (0.021) | (0.028) | (0.010) | (0.017) | (0.011) |
| Full subsidy | -0.001 | -0.028 | -0.025 | -0.010 | -0.049* | -0.020 |
| | (0.020) | (0.044) | (0.019) | (0.013) | (0.025) | (0.013) |
| R-squared | 0.121 | 0.137 | 0.143 | 0.114 | 0.141 | 0.086 |
| **Panel C** | | | | | | |
| 1/3 subsidy | -0.019 | -0.014 | -0.027 | 0.016 | -0.021 | -0.002 |
| | (0.020) | (0.023) | (0.038) | (0.012) | (0.016) | (0.013) |
| 2/3 subsidy | -0.006 | 0.001 | 0.014 | 0.013 | -0.003 | 0.003 |
| | (0.015) | (0.030) | (0.032) | (0.011) | (0.024) | (0.013) |
| Full subsidy | -0.001 | -0.028 | -0.025 | -0.010 | -0.049* | -0.020 |
| | (0.019) | (0.044) | (0.018) | (0.013) | (0.025) | (0.013) |
| R-squared | 0.121 | 0.138 | 0.144 | 0.114 | 0.141 | 0.086 |
| Control group mean | 0.038 | 0.101 | 0.033 | 0.018 | 0.046 | -0.011 |
| Number of observations | 1,200 | 1,566 | 1,196 | 1,263 | 1,622 | 6,191 |

Notes: This table corresponds to Table 5, but the sample is restricted to subsidy only and control groups. All regressions include a standard set of covariates (individual, household, and community) and baseline measure of dependent variable. Robust standard errors clustered at community level are reported in parentheses. *, **, and *** denote statistical significance at 10 %, 5 %, and 1 % levels, respectively.

Table B.4: Effects on Health Care Utilization with Restricted Sample (Long Run)

| | Long run | | | | | |
|---|---|---|---|---|---|---|
| | Visited health facility in last four weeks | Visited health facility in last six months | # of visits in last four weeks | Visited Facility for malaria treatment in the last four weeks | Made an out-of-pocket for health service in the last six months | Standardized treatment effects |
| | (1) | (2) | (3) | (4) | (5) | (6) |
| **Panel A** | | | | | | |
| Any subsidy | 0.041*** | 0.096*** | 0.031** | 0.028* | 0.003 | 0.039*** |
| | (0.013) | (0.026) | (0.012) | (0.014) | (0.006) | (0.010) |
| R-squared | 0.109 | 0.121 | 0.106 | 0.104 | 0.090 | 0.092 |
| **Panel B** | | | | | | |
| Partial subsidy | 0.045*** | 0.086*** | 0.033** | 0.030** | 0.005 | 0.041*** |
| | (0.013) | (0.020) | (0.012) | (0.014) | (0.007) | (0.010) |
| Full subsidy | 0.020 | 0.146** | 0.024 | 0.020 | -0.006 | 0.029 |
| | (0.018) | (0.060) | (0.020) | (0.019) | (0.009) | (0.019) |
| R-squared | 0.110 | 0.123 | 0.106 | 0.104 | 0.091 | 0.092 |
| **Panel C** | | | | | | |
| 1/3 subsidy | 0.012 | 0.071*** | 0.012 | 0.013 | -0.001 | 0.025** |
| | (0.010) | (0.025) | (0.009) | (0.009) | (0.009) | (0.007) |
| 2/3 subsidy | 0.070*** | 0.097*** | 0.048** | 0.043* | 0.009 | 0.053*** |
| | (0.018) | (0.032) | (0.019) | (0.021) | (0.009) | (0.017) |
| Full subsidy | 0.020 | 0.146** | 0.024 | 0.020 | -0.006 | 0.030 |
| | (0.019) | (0.060) | (0.020) | (0.020) | (0.009) | (0.019) |
| R-squared | 0.117 | 0.123 | 0.110 | 0.107 | 0.091 | 0.094 |
| Control group mean | 0.014 | 0.044 | 0.011 | 0.009 | 0.012 | -0.021 |
| Number of observations | 1,236 | 1,546 | 1,238 | 1,236 | 1,546 | 6,180 |

Notes: This table corresponds to Table 6, but the sample is restricted to subsidy only and control groups. All regressions include a standard set of covariates (individual, household, and community) and baseline measure of dependent variable. Robust standard errors clustered at community level are reported in parentheses. *, **, and *** denote statistical significance at 10 %, 5 %, and 1 % levels, respectively.

Table B.5: Effects on Health Status with Restricted Sample (Short Run)

| | Short run | | | | |
|---|---|---|---|---|---|
| | Healthy or very healthy | # Days ill last four weeks | Could not perform normal daily activities due to illness last four weeks | # days could not perform normal daily activities due to illness in the last four weeks | Standardized treatment effects |
| | (1) | (2) | (3) | (4) | (5) |
| **Panel A** | | | | | |
| Any subsidy | 0.148*** | -0.421** | -0.025 | -0.315 | -0.037** |
| | (0.043) | (0.185) | (0.017) | (0.447) | (0.017) |
| R-squared | 0.346 | 0.139 | 0.136 | 0.141 | 0.107 |
| **Panel B** | | | | | |
| Partial subsidy (positive price) | 0.152*** | -0.445** | -0.021 | -0.203 | -0.033* |
| | (0.044) | (0.196) | (0.019) | (0.497) | (0.018) |
| Full subsidy (free) | 0.130* | -0.308 | -0.041 | -0.854 | -0.058** |
| | (0.076) | (0.283) | (0.038) | (0.619) | (0.027) |
| R-squared | 0.346 | 0.139 | 0.136 | 0.142 | 0.107 |
| **Panel C** | | | | | |
| 1/3 subsidy | 0.157*** | -0.740** | -0.044 | -0.916 | -0.061** |
| | (0.047) | (0.300) | (0.031) | (0.728) | (0.027) |
| 2/3 subsidy | 0.147** | -0.225 | -0.005 | 0.343 | -0.011 |
| | (0.061) | (0.250) | (0.021) | (0.517) | (0.021) |
| Full subsidy (free) | 0.130 | -0.298 | -0.040 | -0.836 | -0.057** |
| | (0.077) | (0.287) | (0.038) | (0.617) | (0.027) |
| R-squared | 0.346 | 0.141 | 0.137 | 0.145 | 0.109 |
| Control group mean | 0.818 | 0.616 | 0.082 | 1.376 | 0.011 |
| Number of observations | 478 | 1,597 | 1,603 | 1,549 | 5,081 |

Notes: This table corresponds to Table A11, but the sample is restricted to subsidy only and control groups. All regressions include a standard set of covariates (individual, household, and community) and baseline measure of dependent variable. Robust standard errors clustered at community level are reported in parentheses. *, **, and *** denote statistical significance at 10 %, 5 %, and 1 % levels, respectively.

Table B.6: Effects on Health Status with Restricted Sample (Long Run)

| | Long run | | | | |
|---|---|---|---|---|---|
| | Healthy or very healthy | # Days ill last four weeks | Could not perform normal daily activities due to illness last four weeks | # days could not perform normal daily activities due to illness in the last four weeks | Standardized treatment effects |
| | (1) | (2) | (3) | (4) | (5) |
| **Panel A** | | | | | |
| Any subsidy | -0.128** | 0.249** | 0.042*** | 0.243** | 0.050*** |
| | (0.047) | (0.097) | (0.013) | (0.090) | (0.014) |
| R-squared | 0.416 | 0.095 | 0.132 | 0.094 | 0.083 |
| **Panel B** | | | | | |
| Partial subsidy | -0.133** | 0.296*** | 0.045*** | 0.268** | 0.055*** |
| | (0.058) | (0.095) | (0.013) | (0.097) | (0.014) |
| Full subsidy | -0.111 | 0.019 | 0.027* | 0.122 | 0.026 |
| | (0.126) | (0.157) | (0.013) | (0.097) | (0.020) |
| R-squared | 0.416 | 0.096 | 0.133 | 0.095 | 0.084 |
| **Panel C** | | | | | |
| 1/3 subsidy | -0.115** | 0.287** | 0.027*** | 0.253** | 0.045*** |
| | (0.051) | (0.107) | (0.010) | (0.109) | (0.014) |
| 2/3 subsidy | -0.149 | 0.303** | 0.059*** | 0.279** | 0.063*** |
| | (0.088) | (0.128) | (0.018) | (0.123) | (0.019) |
| Full subsidy | -0.110 | 0.019 | 0.027* | 0.123 | 0.026 |
| | (0.127) | (0.157) | (0.013) | (0.098) | (0.020) |
| R-squared | 0.416 | 0.096 | 0.135 | 0.095 | 0.084 |
| Control group mean | 0.792 | 0.355 | 0.012 | 0.083 | -0.019 |
| Number of observations | 416 | 1,531 | 1,530 | 1,475 | 4,814 |

Notes: This table corresponds to Table A12, but the sample is restricted to subsidy only and control groups. All regressions include a standard set of covariates (individual, household, and community) and baseline measure of dependent variable. Robust standard errors clustered at community level are reported in parentheses. *, **, and *** denote statistical significance at 10 %, 5 %, and 1 % levels, respectively.